\documentclass[preprint,preprintnumbers,superscriptaddress,showkeys,showpacs,nofootinbib,aps]{revtex4-1}

\usepackage{graphicx}

\usepackage{amssymb}
\usepackage{amsmath}
\usepackage{mathrsfs}
\usepackage{dsfont}

\usepackage{color}

\usepackage{slashed}
\usepackage{amscd}
\usepackage{amssymb} 
\usepackage{latexsym}
\usepackage{verbatim}
\usepackage{mathbbol}
\usepackage{hyperref}
\usepackage{bbold}
\newcommand{\Tr}[1]{\mathrm{Tr}\left(#1\right)}

\newcommand{\brackets}[1]{\langle #1 \rangle}

\newcommand{\rbrackets}[1]{\left( #1 \right)}

\newcommand{\epow}[1]{\mathrm{e}^{#1}}

\newcommand{\gammafive}{\gamma_{5}}

\newcommand{\mps}{m_{PS}}

\newcommand{\fermi}{\,\mathrm{fm}}

\newcommand{\mev}{\,\mathrm{MeV}}

\newcommand{\ubar}{\bar{u}}
\newcommand{\dbar}{\bar{d}}

\newcommand{\Jbar}{\bar{J}}

\newcommand{\balign}{\begin{align}}
\newcommand{\ealign}{\end{align}}
\newcommand{\beq}{\begin{equation}}
\newcommand{\eeq}{\end{equation}}
\newcommand{\balignat}[1]{\begin{alignat}{#1}}
\newcommand{\ealignat}{\end{alignat}}

\newcommand{\bfig}{\begin{figure}}
\newcommand{\efig}{\end{figure}}

\newcommand{\bc}{\begin{center}}
\newcommand{\ec}{\end{center}}

\newcommand{\btab}{\begin{table}}
\newcommand{\etab}{\end{table}}

\newcommand{\bcom}{\begin{comment}}
\newcommand{\ecom}{\end{comment}}

\newcommand{\bitem}{\begin{itemize}}
\newcommand{\eitem}{\end{itemize}}

\newcommand{\benum}{\begin{enumerate}}
\newcommand{\eenum}{\end{enumerate}}

\newcommand{\pvec}{\vec{p}}
\newcommand{\Pvec}{\vec{P}}
\newcommand{\qvec}{\vec{q}}

\newcommand{\xvec}{\vec{x}}

\newcommand{\yvec}{\vec{y}}

\newcommand{\zvec}{\vec{z}}
\newcommand{\kvec}{\vec{k}}

\newcommand{\evec}{\vec{e}}

\newcommand{\Llagrange}{\mathcal{L}}

\newcommand{\matelem}{\mathcal{M}}

\newcommand{\refeq}[1]{(\ref{#1})}
\newcommand{\reffig}[1]{[\ref{#1}]}
\newcommand{\reftab}[1]{\{\ref{#1}\}}

\newcommand{\transferMatrix}{\hat{\mathrm{T}}}
\newcommand{\Hamiltonian}{\hat{\mathrm{H}}}
\newcommand{\chisqrPerDof}{\chi^2/\mathrm{dof}}

\def\presuper#1#2%
{\mathop{%
  \mathopen{\vphantom{#2}}^{#1}%
  \kern-\scriptspace%
  #2}
}

\def\MIT{Center for Theoretical Physics, Laboratory for Nuclear Science and Department of Physics, Massachusetts Institute of Technology, Cambridge, Massachusetts 02139, U.S.A}
\def\RFWUB{Helmholtz-Institut fur Strahlen- und Kernphysik, Rheinische Friedrich-Wilhelms-Universit\"at Bonn, Nu{\ss}allee 14-16, D-53115 Bonn, Germany}
\def\CYI{Computation-based Science and Technology Research Center, Cyprus Institute, 20 Kavafi Str., 2121 Nicosia, Cyprus}
\def\UCY{Department of Physics, University of Cyprus, P.O. Box 20537, 1678 Nicosia, Cyprus}
\def\RIKEN{RIKEN BNL Research Center, Brookhaven National Laboratory, Upton, NY 11973, USA}

\begin{document}

%%%\preprint{}
\author{C.~Alexandrou}
\affiliation{\UCY}
\affiliation{\CYI}

\author{J.~W.~Negele}
\affiliation{\MIT}

\author{M.~Petschlies}
\affiliation{\CYI}
\affiliation{\RFWUB}

\author{A.~V.~Pochinsky}
\affiliation{\MIT}

\author{S.~N.~Syritsyn}
\affiliation{\RIKEN}

\title{Study of decuplet baryon resonances from lattice QCD}


\begin{abstract}
A lattice QCD study of the strong decay width and coupling constant of decuplet baryons to an octet baryon - pion state is presented.
The transfer matrix method is used to obtain the overlap of lattice states with decuplet baryon quantum numbers on the one hand 
and octet baryon-pion quantum numbers on the other as an approximation to the matrix element of the corresponding transition. 
By making use of leading order effective field
theory, the coupling constants, as well as the widths for the various decay channels are determined. The transitions studied are
$ \Delta \to \pi \,N$,
$\Sigma^* \to \Lambda\,\pi$, $\Sigma^* \to \Sigma\,\pi$ and $\Xi^* \to \Xi\,\pi$.
We obtain results for two ensembles of $N_f=2+1$ dynamical fermion configurations,
one using domain wall valence quarks on a staggered sea at a pion mass of $350\mev$ and
a box size of $3.4\fermi$ and a second one using domain wall sea and valence quarks at pion mass $180 \mev$ and box size $4.5\fermi$.

\end{abstract}

\maketitle

\section{Introduction}

%The study of  resonances 
%starting directly from the fundamental theory of the
%of strong interaction in its discretized Euclidean formulation, lattice  Quantum Chromodynamics (QCD), has  seen a lot of progress.
The study of resonances from first principles using lattice Quantum Chromodynamics (QCD), has progressed significantly.
Most of these studies 
are based on the L\"uscher approach \cite{Luscher:1985dn,Luscher:1986pf}
and extension thereof~\cite{Rummukainen:1995vs,Fu:2011xz,Gockeler:2012yj,Leskovec:2012gb,Doring:2012eu}
%that probes the change of the hadron spectrum by imposing quantization conditions in finite volume for scattering lengths and phase shifts.
that extract scattering lengths and phase shifts from discrete energy levels in a finite volume.
The approach has been generalized to the case of coupled channels~\cite{He:2005ey,Bernard:2008ax,Doring:2011nd,Lang:2011mn,Briceno:2012yi,Hansen:2012tf,Guo:2012hv,Li:2012bi,Briceno:2013bda,Wu:2014vma} and three identical boson scattering~\cite{Briceno:2012rv}, and
a growing number of studies is being carried out in the meson
sector. A pioneering study of  meson-baryon and baryon-baryon scattering lengths
was already carried out twenty years ago~\cite{Fukugita:1994ve} and more recent studies include those by members of the NPLQCD~\cite{Briceno:2013hya,Torok:2009dg} and HALQCD~\cite{Aoki:2011ep} collaborations and other groups~\cite{Lage:2009zv}. Despite this progress, the  application of the L\"uscher
approach  to baryon resonances has been limited since the
method requires very precise data for multiple spatial volumes or various reference frames of different total linear momentum   making
 it computationally
very demanding.

%%% \textcolor{blue}{
Another method to study hadronic resonant decays from lattice QCD 
was proposed in Refs.~\cite{McNeile:2002az,McNeile:2002fh} 
and successfully applied in the study of meson decays~\cite{Michael:2006hf,Michael:2005kw}.
A first application of this transfer matrix method to baryons
was carried out in Refs.~\cite{Alexandrou:2013ata,Alexandrou:2014qka}. The transfer matrix method as applied here
allows 
to extract the width of a resonant hadronic decay, if the resonance width is
small as compared to the resonant energy and well-isolated from other decay channels.
 In such a situation the method allows us to extract
the width  from one kinematic point and it thus provides
currently a computationally feasible calculation of the width in the baryon sector. 
This calculation can be seen as a first attempt to compute the width of an unstable baryon
that allows us to learn about two-particle interpolating fields in the baryon sector and the associated technicalities
and gauge noise.
%%% This
%%% calculation also provides a test-bed for some of   elements for a follow-up calculation using the
%%% L\"uscher method.
%%% }

\par

\begin{figure}[htb]
\centering
\includegraphics[width=0.7\textwidth]{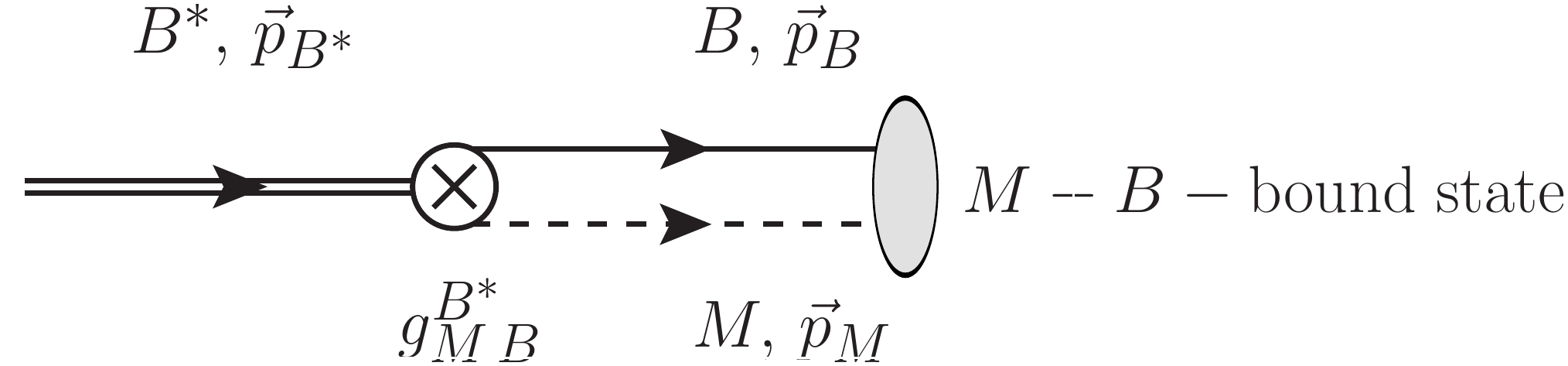}
\caption{Diagram for the transition $B^* \leftrightarrow M B$}
\label{fig:1}
\end{figure}

In this approach, one considers a purely hadronic decay of a  baryon $B^*$ to 
a two-particle  state. In the cases considered in this work,
the two-particle state will be a meson $M$ and a baryon $B$, as illustrated diagrammatically in Fig. \reffig{fig:1}.

%\textcolor{blue}{
We associate a vertex with the tree-level transition graph in Fig.~\reffig{fig:1} and the strength of the interaction at the vertex
is measured in terms of an effective coupling constant $g^{B^*}_{M B}$.
% In an effective field theory with 
%$M,B$ and $B^*$ degrees of freedom, the tree-level diagram in Fig. \reffig
%{fig:1} 
We define this coupling to coincide with the coupling that appears in the leading order continuum effective field theory for
the interaction term of the hadronic fields $M,\,B$ and $B^*$ in the effective Lagrangian.
This will be made more explicit later on in connection with Eq.~ \refeq{eq:effective_lagrangian}.

%%% {\bf Dina: I think we should move this to later one} %Its meaning is equivalent to one of $h_A$ in the Chiral effective field theory approach in Ref. \cite{Pascalutsa:2005vq} by
%the factor of nucleon mass over pion decay constant, $ g^{B^*}_{M B} = h_A\cdot m_N / f_\pi$ .}\\

%%% {\bf Dina: I modified the following}
In order to study a decay $B^* \to M\,B$ in the Euclidean quantum field theory
we need to formulate it in terms of energies or hadronic matrix elements.
In lattice QCD we use interpolating fields to create  states with the quantum of the decuplet baryon
$B^*$ and  the octet baryon $B$ and the meson $M$. 
%Correlation functions in Euclidean space will be either 
%purely real or imaginary. 
We restrict our consideration to the two lowest-lying states with the desired quantum numbers,
which we label by $|\,B^*\rangle $ and $|\,MB\rangle$. If $B^*$ does not decay then its  overlap with $|\,MB\rangle$ is zero and it is an asymptotic state of the theory. These states can then be thought off as the eigenstates of  a non-interacting lattice transfer matrix $\transferMatrix^0$ defined by a Hamiltonian
$\Hamiltonian_0=|B^*\rangle\langle B^*|+|\,MB\rangle\langle\,MB|$.
Our approach here is to study the  
overlap of the states created by the interpolating fields $B^*$ and $M\,B$ for the case where the energy levels  of  these states  are near-degenerate.
 The interaction Hamiltonian to leading order in the
perturbation will then be given by $\Hamiltonian= \Hamiltonian_0
+|B^*\rangle\langle MB|+|\,MB\rangle\langle\,B^*|$. 

%linear combinations of the non-interacting states
%\begin{align}
%  |\,B^*\rangle &\propto |\,B^*\rangle^0 + \alpha\,|\,MB\rangle^0 \,,\quad
%  |\,MB\rangle \propto \alpha^*\,|\,B^*\rangle^0 + |\,MB\rangle^0
% \label{eq:perturbed_states}
%\end{align}
The off-diagonal elements of the Hamiltonian will be  the overlap $\brackets{B^*\,|\,\Hamiltonian \,|\,MB}$. Thus our assumption is
that  a state $|\,B^*\rangle$ created initially at time $t_i$  propagates 
in Euclidean time on the lattice to final time $t_f$, makes one transition
to $ |\,MB\rangle$ 
at any intermediate time step $t \to t + a$ on the lattice.
If the (real valued) lattice transition amplitude is small in terms of the inverse propagation time
$\left( t_f - t_i \right)^{-1}$ and if the energy gap between the states $|\,B^*\rangle$ and $|\,MB\rangle$ is sufficiently small then one can evaluate
the overlap and relate it to the coupling constant and then to the decay width~\cite{McNeile:2002fh}.
We stress that the propagator of the state $B^*$ is fully dressed and so is the propagator of the $M\,B$ state including interactions between the two particles.
A tree-level effective interaction Lagrangian can be written in terms of the fields $B^*, B$ and $M$~\cite{Pascalutsa:2002pi} with 
coupling constant $g^{B^*}_{MB}$, which is related to the overlap $\brackets{B^*\,|\,\Hamiltonian \,|\,MB}$ as will be discussed in section II.

 In order compute the overlap of these states we need to choose  ensembles for which  the energy gap $\delta = E_{B^*} - E_{M B}$ is small in units of the inverse propagation time from initial to final state $\delta \ll 1 / (t_f - t_i)$. Since on a finite lattice the allowed momenta are discretized  the energies will not in general match. Thus, this condition will be only approximately satisfied for the ensembles we have at our disposal.
%This condition ensures the dominance of the characteristic time dependence of the lattice data, from
%which the transition amplitude is extracted.
A second condition that is required  is that the propagation time $t_f-t_i$
is sufficiently large compared to the energy difference between the ground state energy of  $B^*$ and its first excited state as well as between the lowest energy of the $B\,M$ system and its excited state with the same quantum numbers so that only
  the two lowest-lying states of interest
dominate in the transition matrix element.
The extraction of the overlap from lattice measurements is detailed in section II. The transition matrix element $\matelem(B^* \to M B) \propto \brackets{B^* \,|\,H\,|\,M B}$
for the situation in which the energy levels of the two states are degenerate.
Using Fermi's Golden rule one can relate this decay matrix element to the decay width
\begin{align}
  \Gamma^{B^*}_{M\,B} &= 2\pi\,|\matelem(B^* \to M B)|^2\,\rho\,,
  \label{eq:fermis_golden_rule}
\end{align}
where $\rho$ is the density of states at the transition energy.
As already mentioned, in this study we work to  lowest order 
considering only  a single transition amplitude and allowing for
large enough time separation $t_f-t_i$ so only the  lowest states in the initial and final states give the dominating contribution.
To this order we also neglect further elastic rescattering of 
$M\,B$ in the final $MB$ state.
%%% {\bf Dina: I am not sure I understand why the density is related to beglecting the interactions between B and M}} %thus within this appoximation consistent determine the density and normalization of states for asymptotic, infinite volume states.}

In our first study~\cite{Alexandrou:2013ata},
we successfully applied  this approach to study the $\Delta$ resonance using  a hybrid action with
 domain wall valence quarks  on a staggered sea. Here we extend our study to include the decuplet baryons $\Sigma^*$ and $\Xi^*$. In addition, 
we investigate 
 the applicability of  All Mode Averaging (AMA)~\cite{Blum:2012uh} in improving the statistical accuracy  using  the $\Delta$ resonance as test  case.
In this work we also %an
 analyze 
an ensemble of domain wall fermions (DWF) corresponding to a  pion mass of 180~MeV\cite{Arthur:2012opa} for which the energy matching,
in particular for the $\Delta$, is
very well satisfied. The results based 
on this ensemble of $N_f=2+1$ DWF for the widths of the $\Delta$, the $\Sigma^*$ and the $\Xi^*$
as well as the results using the hybrid ensemble for the $\Sigma^*$ and $\Xi^*$
constitute the first determination of the decay widths of these resonances using lattice QCD.

The paper is organized as follows: In section~\ref{sec:technical_details_of_the_method} we present the method and give the technical details.
In section~\ref{sec:Numerical results} we show our lattice QCD results, in section~\ref{sec:Discussion} we discuss these results
and their relation to the decay widths and in section~\ref{sec:Conclusions} we present our conclusions.

\section{Technical details of the method}
\label{sec:technical_details_of_the_method}
The  method that we consider in this work was first described in Refs.~\cite{McNeile:2002az,McNeile:2002fh,McNeile:2004rf}  where it was applied to the study of meson decays. The method 
was extended for the case of the $\Delta$ resonance and first results
were obtained using an ensemble of domain wall valence quarks on an $N_f=2+1$ staggered sea, which we will refer to as hybrid approach~\cite{Alexandrou:2013ata}.
In this section, we explain the technical steps involved 
paying particular attention to the description of the decays of decuplet baryons, which is the focus of this work.

\subsection{Lattice correlation function and normalization}
\label{subsec:Lattice_correlation_function_and_normalization}
We consider the following strong decays of a decuplet baryon to a meson-baryon final state:
\begin{align}
  \Delta   &\to \pi\, N \nonumber\\
  \Sigma^* &\to \pi\, \Sigma\,,\: \pi \,\Lambda \nonumber\\
  \Xi^*    &\to \pi \,\Xi\,,
\label{eq:decuplet_decays}
\end{align}
generically denoted by  $B^* \to M\,B$.
Due to the isospin symmetry of the lattice action, we can choose any isospin channel for each case. In this study, we consider the $\Delta^{++}$, the
$\Sigma^{*+}$ and the $\Xi^{*-}$ with  interpolating fields  given by
\begin{align}
  J_{B^*}^{\mu\alpha}(t,\Pvec) &= \sum\limits_{\xvec}\,J_{B^*}^{\mu\alpha}(t,\xvec)\,\epow{i\Pvec\xvec} \nonumber\\
  J_{\Delta^{++}}^{\mu\alpha}(x) &= \epsilon_{abc}\,\rbrackets{u^{a T}(x)\,C\,\gamma^\mu\,u^{b}(x)}\,u^{\alpha\,c}(x)\nonumber\\
  J_{\Sigma^{*+}}^{\mu\alpha}(x) &= \frac{1}{\sqrt{3}}\,\epsilon_{abc}\,\Big[
  \rbrackets{u^{a T}(x)\,C\,\gamma^\mu\,u^b(x)}\,s^{\alpha\,c}(x) 
   + 2\,\rbrackets{s^{a T}(x)\,C\,\gamma^\mu\,u^b(x)}\,u^{\alpha\,c}(x) \Big] \nonumber\\
  J_{\Xi^{*-}}^{\mu\alpha}(x) &= \epsilon_{abc}\,\rbrackets{s^{a T}(x)\,C\,\gamma^\mu\,d^b(x)}\,s^{\alpha\,c}(x) \,,
  \label{eq:Bstar_interpolating_fields}
\end{align}
where  lower case Latin (Greek) letters denote color (spin) indices and $C = i\gamma_0\,\gamma_2$
is the charge conjugation matrix.

As interpolating fields for the meson-baryon states we take the   product of the interpolating fields of  the corresponding meson and baryon:

\begin{align}
  J_{M B}^\alpha(t,\pvec_M,\pvec_B) &= \sum\limits_{\yvec,\zvec}\,J_{M}(t,\yvec)\,J_B^\alpha(t,\zvec)\,\epow{i( \pvec_M \yvec + \pvec_B \zvec)}\nonumber\\
  J_{\pi^+}(y) &= \dbar(y)\,\gammafive\,u(y)\nonumber\\
  J_{N}^\alpha(z) &=  \epsilon_{abc}\,\rbrackets{u^{a T}(z)\,C\,\gammafive\,d^b(z)}\,u^{\alpha\,c}(z)\nonumber\\
  J_{\Lambda}^\alpha(z) &=\frac{1}{\sqrt{6}}\,\epsilon_{abc}\,\Big[
    2\,\rbrackets{u^{a T}(z)\,C\,\gammafive\,d^b(z)}\,s^{\alpha\,c}(z) \nonumber\\
  &\quad + \rbrackets{u^{a T}(z)\,C\,\gammafive\,s^b(z)}\,d^{\alpha\,c}(z)
    - \rbrackets{d^{a T}(z)\,C\,\gammafive\,s^b(z)}\,u^{\alpha\,c}(z)
\Big]\nonumber\\
J_{\Sigma^0}^{\alpha} (z) &= \frac{1}{\sqrt{2}}\,\epsilon_{abc}\,\Big[
  \rbrackets{u^{a T}(z)\,C\,\gammafive\,s^b(z)}\,d^{\alpha\,c}(z)
  + \rbrackets{d^{a T}(z)\,C\,\gammafive\,s^b(z)}\,u^{\alpha\,c}(z) \Big]\nonumber\\
J_{\Xi^0} &= \epsilon_{abc}\,\rbrackets{s^{T a}(z)\,C\,\gammafive\,u^b(z)}\,s^{\alpha\,c}(z) \,.
\label{eq:MB_interpolating_fields}
\end{align}
Previous studies have shown that the two-hadron interpolating fields given in Eq.~\refeq{eq:MB_interpolating_fields}
will primarily overlap with two-hadron states~\cite{Dudek:2012xn},  while the
single-hadron interpolating fields in Eq.~\refeq{eq:Bstar_interpolating_fields} will have a dominant overlap with single-hadron states.

In the following, we consider kinematics where  the total momentum is zero,
so in the first line of Eq.~\refeq{eq:Bstar_interpolating_fields} we set $\Pvec=0$
and $\pvec_M = -\pvec_B$ in Eq.~\refeq{eq:MB_interpolating_fields}.
Using these interpolating fields we build the two-point correlation functions for $C_{B^*-B^*}$, $C_{B^*-M B}$ and $C_{M B-M B}$ as follows:
\begin{align}
  C_{B^*-B^*}(t_f-t_i) &= \Tr{
    \frac{1}{4}(\mathbb{1}+\gamma_0)\,P^{3/2}_{k l}\,\sum\limits_{\xvec}\,\brackets{J^{l}_{B^*}(t_f,\xvec)\,\Jbar_{B^*}^{k}(t_i,\zvec)}
  }\label{eq:C_Bstar_Bstar}\\
  C_{B^*-M B}^k(t_f-t_i,\qvec) &= \Tr{
    \frac{1}{4}(\mathbb{1}+\gamma_0)\,P^{3/2}_{k l}\,\sum\limits_{\xvec,\yvec}\,\brackets{J^{l}_{B^*}(t_f,\xvec)\,J^\dagger_M(t_i,\yvec)\,\Jbar_{B}(t_i,\zvec)}
  }\,\epow{-i\qvec(\yvec-\zvec)}
  \label{eq:C_Bstar_MB}\\
  C_{M B - M B}(t_f-t_i,\kvec_f,\kvec_i) &= \Tr{
    \frac{1}{4}(\mathbb{1}+\gamma_0)\,\sum\limits_{\xvec,\yvec,\yvec'}\,\brackets{J_{B}(t_f,\xvec)\,J_M(t_f,\yvec)\,J^\dagger_M(t_i,\yvec')\,\Jbar_{B}(t_i,\zvec)}\,
  }\nonumber\\
    &\qquad \times \epow{i\kvec_f(\yvec-\xvec) - i\kvec_i(\yvec'-\zvec)}
  \label{eq:C_MB_MB}
\end{align}
with a fixed source location $(t_i,\,\zvec)$.
All correlators are defined to include a parity projection 
$\frac{1}{4}\,\rbrackets{  \mathbb{1} \pm \gamma_0 }$.
In addition, $C_{B^*-B^*}$ and $C_{B^*-M B}$ include a projector to spin 3/2, which at zero total momentum is given by
\begin{align}
  P^{3/2}_{i k} &= \delta_{i k}\,\mathbb{1} - \frac{1}{3}\,\gamma_i\,\gamma_k\,,\quad i,k=1,2,3\,.
  \label{eq:spin32_projector}
\end{align}
%%%The interpolating fields generate an overlap with the desired on-shell states, which is parametrized as
%%%\begin{align}
%%%	\brackets{B^*, \Pvec'\,|\,\Jbar_{B^*} (t,\Pvec=0) \,|\,0} &= Z_{B^*}(0)\,\ubar_{B^*}(\Pvec)\,
%%%	\epow{E_{B^*}(\Pvec=0) t}\,V\,\delta_{\Pvec' 0} \nonumber\\
%%%	\brackets{M B, \qvec'\,|\,\Jbar_{M B} (t,\qvec) \,|\,0} &= Z_{M B}(\Pvec=0,\qvec^2)\,\ubar_{B}(-\qvec)\,\epow{E_{M B}(\Pvec=0, \qvec) t }\,
%%%	V\,\delta_{\qvec \qvec'} \,.
%%%	\label{eq:invertpolating_fields_overlap}
%%%\end{align}

To cancel the unknown overlaps of the interpolating fields with the states
 we construct the ratio
\begin{align}
	R^{B^*}_{M B,\,k}\left( t_f-t_i,\,\qvec,\,\kvec_f,\,\kvec_i\right) &= \frac{
		C_{B^*-M B}^k(t_f-t_i,\qvec)
	}{ \sqrt{ C_{B^*-B^*}(t_f-t_i) \times C_{M B - M B}(t_f-t_i,\kvec_f,\kvec_i) }
	}\,.
	\label{eq:ratio_definition}
\end{align}
In this work we always consider the case $\kvec_f = \kvec_i$ and $|\qvec| = |\kvec_f|$. The 2-point function $C_{M B - M B}$
then only depends on $|\kvec_f|$ and the ratio can be characterized by a single vector $\kvec = \qvec$ as 
$R^{B^*}_{M B,\,k} = R^{B^*}_{M B,\,k}(t_f-t_i,\kvec)$\\

\paragraph{Alignment and polarization for $B^* - M B$, momentum averages and angular momentum:} 
Within leading order in  effective field theory, a non-vanishing signal in $C_{B^* - M B}^k(t,\qvec)$ only arises when the relative momentum vector is aligned 
or anti-aligned with the spin projection appearing  in the correlation function, i.e. when $\qvec \cdot \evec_k \ne 0$, where $\evec_k$ denotes the
 unit vector in the $k-$direction.
%%%\footnote{
The vertex for the fields $B^*,\,M,\,B$ in the effective Lagrangian following our notation is given by
\begin{align}
  \Llagrange_{I} \sim g^{B^*}_{MB}\,\bar{B}^*_\mu\,\partial_\mu\,M^a\,T^a\,B
  \label{eq:effective_lagrangian}
\end{align}
with matrices $T^a$, which contain the Clebsch-Gordan coefficients for coupling isospin channels.
\begin{align*}
  \left( T^a \right)_{ik} &=  \brackets{ I_{B^*}=3/2,\,I_{B^*}^3= i\,|\,I_{B}=1/2,\,I_{B}^3=k;\,I_{M}=1,\,I_{M}^3=a }\,,\\
  i\,&\in\,\{-3/2,\,-1/2,\,+1/2,\,+3/2\}\,,\quad k \,\in\,\{-1/2,\,+1/2\}\,,\quad a\,\in\,\{-1,0,1\}\,.
\end{align*}

%%%}
We perform our calculations with one unit of relative momentum, $|\qvec| = 2\pi/L$, such that $\qvec = (\pm 1,0,0)\,2\pi/L$ or a permutation thereof and
thus look at the six combinations $C_{B^*-MB}^k(t, \pm 2\pi/L\,\evec_k),\,k=1,2,3$.\\

The $B^* - M B$ correlator is projected to its spin-3/2 component with $P^{3/2}$. Moreover,
the average over positive and negative momentum effectively means
that for the $M B$ state we use the interpolating field in its center-of-mass frame
\begin{align}
	J_{M B}^i\left(t,\Pvec=0, k=2\pi/L\right)  &= 
	\frac{1}{2} \,
	\sum\limits_{i'=1,2,3}\,P^{3/2}_{i i'}\,
	\left[ J_{M B}\left( t,\Pvec=0, k \evec_{i'} \right) - J_{M B}\left( t,\Pvec=0, -k \evec_{i'} \right) \right]\,.
	\label{eq:averaged_MB_interpolator}
\end{align}
In a partial wave expansion, we find that the dominant state excited by the interpolating field given in  Eq.~\refeq{eq:averaged_MB_interpolator} will have orbital angular $l = 1$. The coupling of the $l=1$  to the
 nucleon  state with spin 1/2 are projected to the component with
 total angular momentum 3/2. Thus the operators in Eq. \refeq{eq:averaged_MB_interpolator} and the projected operators in Eq. \refeq{eq:Bstar_interpolating_fields} transform
 under the spin-3/2 representation of the Lorentz group in the continuum, which is subduced into 
%%% {\bf Dina Check:}
 irreducible representation $\Lambda^P = H^+$ (positive parity) of the double cover $O^D$ of the octahedral group on the lattice (Table III in Ref. \cite{Basak:2005ir}).
The irreducible representation $H$ contains
an overlap with higher partial waves. What simplifies the calculation at hand, is that we only consider the ground state at large Euclidean
time, in which all channels but the desired one with $l = 1, J^P = 3/2^+$ are exponentially suppressed.

All the standard spin-3/2 interpolating fields  $J_{B^*}$ involve the spin structure $(q^T\,C\,\gamma_k\,q)\,q$ and the parity operation in the center-of-mass
frame acts as
\begin{align}
	\mathrm{P} &: \, J_{B^*}^k(t,\Pvec=0) \to \gamma_0\,J^k_{B^*}(t,\Pvec=0)\,.
	\label{eq:j_Bstar_parity}
\end{align}
The spin-3/2 projector, $P^{3/2}$, and the projector to the component with definite parity commute, such that we have the trivial action
\begin{align}
	\mathrm{P} &: \, \frac{1}{4}(1+\gamma_0)\,P^{3/2}_{k k'}\,J_{B^*}^{k'}(t,\Pvec=0) \to \frac{1}{4}(1+\gamma_0)\,P^{3/2}_{k k'}\,J_{B^*}^{k'}(t,\Pvec=0)\,.
	\label{eq:j_Bstar_parity_2}
\end{align}
On the other hand, for the $M B-$state with the pseudoscalar meson field we have in the center-of-mass frame and with relative momentum $\kvec$
\begin{align}
\mathrm{P} &: \, \frac{1}{4}(1+\gamma_0)\,J_{M B}(t,\Pvec=0,\kvec) \to -\frac{1}{4}(1+\gamma_0)\,J_{M B}(t,\Pvec=0,-\kvec)\,.
\label{eq:j_MB_parity}
\end{align}
Since parity is a symmetry of the lattice action and the $B^* - B^*$ and $M B - M B$ two-point functions are even under parity, we expect the following relation to hold for the ratio
\begin{align}
	R^{B^*}_{M B}(t_t-t_i,\kvec) &= - R^{B^*}_{M B}(t_t-t_i,-\kvec)\,.
	\label{eq:ratio_under_parity}
\end{align}

%%% \textbf{We could check to see, that the Delta int. fields are a basis for $H_3$
%%% and that also these $J_{MB}^i$ coupling I guess from lattice $G_? + T_1 \to H_3$, so it is clear, that
%%% we have $J_{B^*}^k\,J_{M B}^l \propto \delta^{kl}$ from group orthogonality relations}

\begin{figure}[htpb]
	\begin{center}
		\includegraphics[width=0.25\textwidth ]{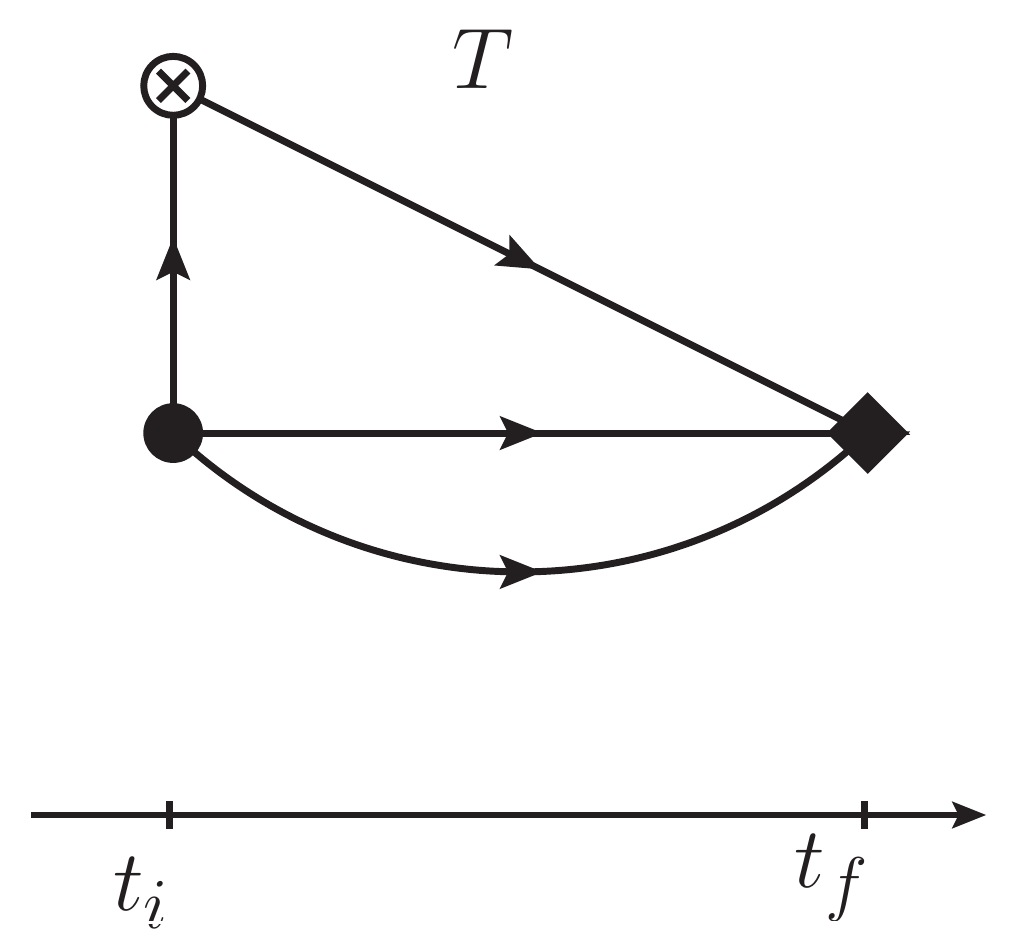}
		\includegraphics[width=0.25\textwidth ]{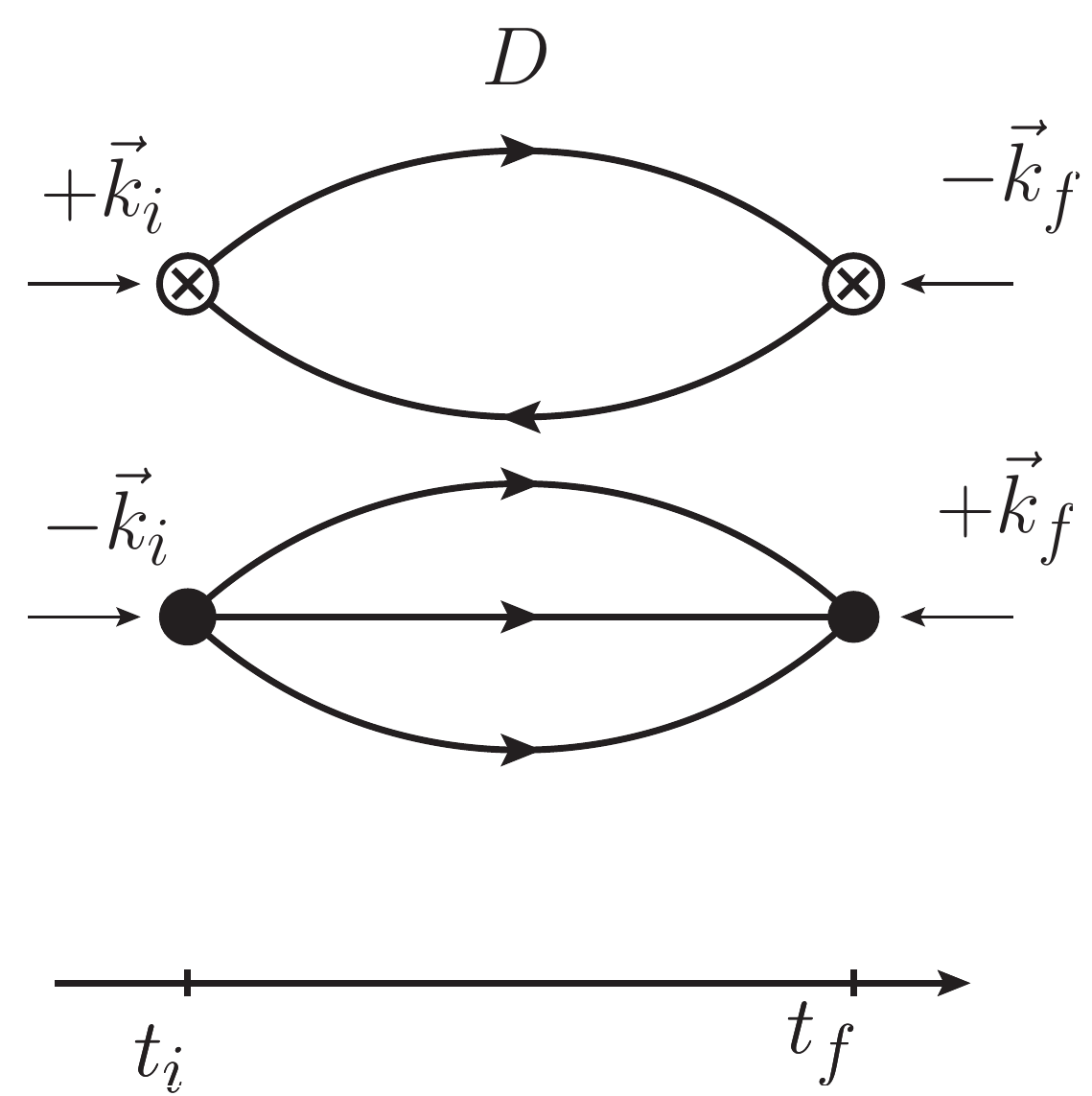}
		\includegraphics[width=0.7\textwidth ]{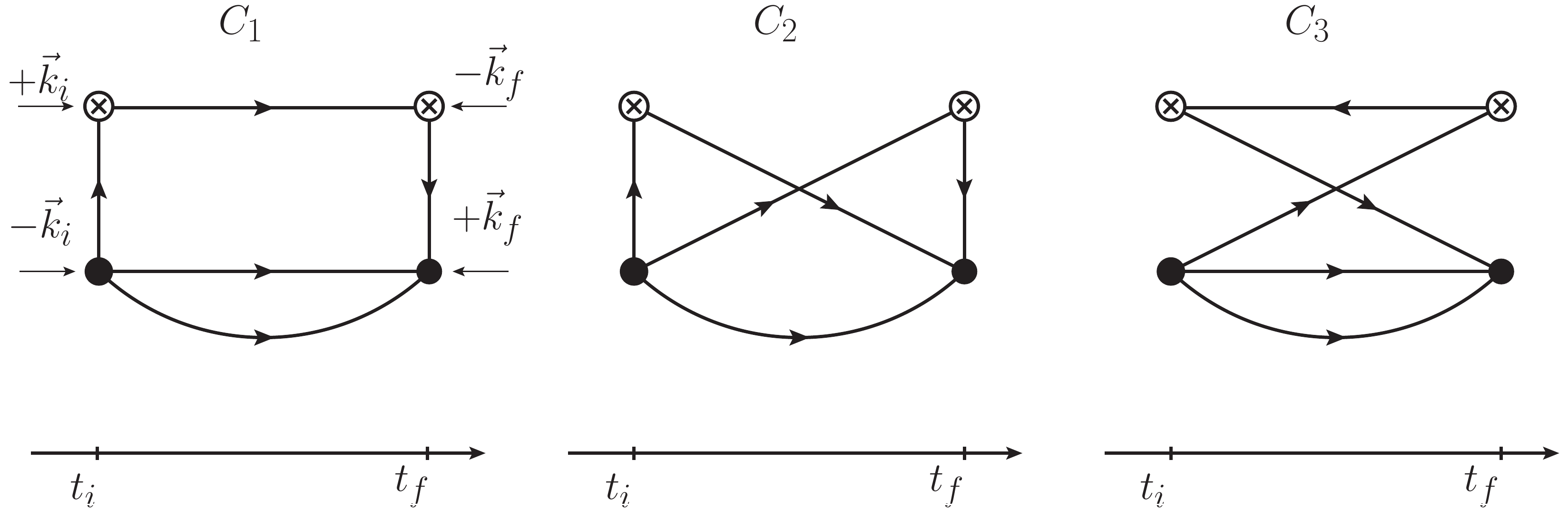}
	\end{center}
	\caption{Diagrams representing different types of Wick contractions for $C_{B^*-MB}$ (diagram $T$) and $C_{MB-MB}$ (upper right: quark-disconnected $D$, lower: quark-connected $C_{1,2,3}$).}
	\label{fig:2}
\end{figure}

\paragraph{Quark-connected and disconnected diagrams:}
The Wick contractions for the correlation function $C_{MB-MB}$ can be represented by two types of diagrams,
as shown in Fig.~\reffig{fig:2}: quark-disconnected ($D$, upper right) and quark-connected (lower diagrams $C_{1,2,3}$).

We make two simplifications: i) We neglect the quark-connected diagrams $C_1$, $C_2$ and $C_3$
and ii) approximate the quark-disconnected diagram by the product of expectation values
of the individual meson and baryon propagators. This results in a significant reduction in the computational cost. The contractions for diagrams $T$ and $D$, as performed in this work,
require one quark propagator and one sequential propagator through the source timeslice $t_i$.
In contrast, the full contractions for the quark-connected diagrams  require all-to-all propagators. We thus set
\begin{align}
	C_{MB-MB}(t_f-t_i,\kvec_f,\kvec_i) &\to C_{MB-MB}(t_f-t_i,\kvec_M,\kvec_B) \nonumber \\
	&= \brackets{C_{M-M}(t_f-t_i,\kvec_M)} \times 
	\Tr{\frac{1}{4}(1+\gamma_0)\, \brackets{C_{B-B}(t_f-t_i,\kvec_B)} },
	\label{eq:C_MB_MB_simplified}
\end{align}
where the meson and baryon propagators are Fourier transformed with independent momenta $\kvec_M$ and $\kvec_B$, respectively. In this form the $M-B$ two-point function
depends only on the squared momenta $\kvec_M^2$ and $\kvec_B^2$ and for $C_{MB-MB}$ we only use combinations with equal modulus such that we can simply
replace the dependence on the pair $(\kvec_M,\kvec_B))$ by a single vector $\kvec$. 

Thus, in the asymptotic region of large time separation $(t_f - t_i) / a \gg 1$, we can express this approximation as
\begin{align}
C_{MB-MB}(t_f-t_i,\kvec_f,\kvec_i) &\overset{ (t_f - t_i)/a\gg 1}{\propto} 
  Z_{MB}(\kvec_f,\kvec_i) \,\epow{-E_{MB}(\kvec_f,\kvec_i)(t_f-t_i)}\,u_{MB}(\kvec_f)\,\ubar_{MB}(\kvec_i) + \ldots \nonumber\\
&\approx  Z_M(\kvec^2)\,\epow{-E_M(\kvec^2)(t_f-t_i)} \, Z_{B}(\kvec^2) \,\epow{-E_{B}(t_f-t_i)}\,u_{B}(\kvec_f)\,\ubar_{B}(\kvec_i) + \ldots
\label{eq:C_MB_MB_asymptotic}
\end{align}

\subsection{Extraction of coupling and the matrix element}
 For all
 baryons $B^*$ to $M\,B$ we restrict the lattice Hilbert space to
two states $|\,B^*\rangle$ and $|\,M B\rangle$ as the dominant baryon and baryon-meson states, respectively~\cite{Alexandrou:2013ata}. In terms of this two-dimensional subspace,
the transfer matrix is parametrized as
\begin{align}
  \transferMatrix &= \epow{-a\bar{E}}\,
  \begin{pmatrix}
    \epow{-a\delta/2} & ax \\ ax & \epow{+a\delta/2}
  \end{pmatrix}\,.
\label{eq:transfer_matrix_subspace}
\end{align}
In accordance with the assumption of a small energy gap we take $\bar{E} = (E_{B^*} + E_{M B}) / 2$ as the mean of the energies of the states and 
$\delta = E_{M B} - E_{B^*}$ as their difference. Fixing an initial and final lattice timeslice $t_i$ and $t_f$ and summing over all possibilities
for a {\it single} transition from $B^*$ to $M B$, which can occur at any intermediate time-slices between $t_i$ and $t_f$, it follows that
\begin{align}
\brackets{B^*,\,t_f\,|\,M B,\,t_i} &= \brackets{B^*\,|\,\epow{-H(t_f-t_i)}\,|\,M B} = \brackets{B^*\,|\,\transferMatrix^{n_{fi}}\,|\,M B} \nonumber\\
&= \sum\limits_{n=0}^{n_{fi}-1}\,\epow{-(\bar{E}-\delta/2)t_n}\,\brackets{B^*\,|\,\transferMatrix\,|\,M B}\,
  \epow{-(\bar{E}+\delta/2)(\Delta t_{fi}-t_n-a)} + \ldots\nonumber\\
&= ax\,\frac{\sinh(\delta\,\Delta t_{fi}/2)}{\sinh(a\delta/2)}\,\epow{-\bar{E}\Delta t_{fi}} + \ldots
\label{eq:derivation_fit_formula}
\end{align}
with $\Delta t_{fi} = t_f - t_i = a n_{fi}$. In Eq.~\refeq{eq:derivation_fit_formula} the ellipsis  denotes contributions
of higher orders in the matrix elements $\brackets{B^*\,|\,\transferMatrix\,|\,M B}$ and
$\brackets{M B\,|\,\transferMatrix\,|\,B^*}$, which are at least quadratic in the time separation $\Delta t_{fi}$.
As a consequence, by extracting the term linear in $\Delta t_{fi}$ we get the transfer matrix element 
\begin{align}
  ax &= \brackets{B^*\,|\,T\,|\,M B} \xrightarrow{a \to 0} -a \brackets{B^*\,|\,\mathrm{H}\,|\,M B}\,.
\label{eq:slope_to_matrix_element}
\end{align}

Given the time-dependence of the overlap in Eq.~\refeq{eq:slope_to_matrix_element},  we use two different fit ans\"atze  given by
\begin{align}
	f_1(t) &= c_0 + c_1\,\sinh\left( c_2\,t/(2a) \right)/(c_2/2),
	\label{eq:fit_type_1}\\
	f_2(t) &= c_0 + c_1\,\frac{t}{a} + c_2 \left( \frac{t}{a} \right)^3 + \ldots
	\label{eq:fit_type_2}
\end{align}
In both cases, we are primarily interested in the parameter $c_1$. Although in principle we could take into account the next terms
denoted by the ellipsis in Eq.~\refeq{eq:fit_type_2}, in practice we will not need to go beyond
$c_2$ to obtain a good fit to the available lattice data with their present accuracy. The parameter $c_0$ allows for an offset at $t=0$,
which can originate from lattice artifacts or
contributions from excited states giving an overlap at zero time, i.e. with no  insertion of  the transfer matrix. Given that
Eq.~\refeq{eq:derivation_fit_formula}
contains a lattice version of the energy Dirac-$\delta$ function for the finite temporal lattice extent,
Eq.~\refeq{eq:fit_type_1} allows for fitting the data taking into account a non-zero energy gap.
Besides taking into account a finite energy gap, 
which will be the only contribution to it if the transition happens once in the
path integral,
the $c_2$-term
may also effectively be including next-to-leading order contributions
arising from overlaps from other intermediate states.

The value of the parameter $c_2$ extracted from fitting
 $f_1$ and the one extracted from fitting
$f_2$ can be related at order $t^3$ after expanding $f_1$. We indeed 
find that these two values  of the $c_2$ parameter show a strong correlation, as expected. Yet in some cases  a different value
is extracted  and hence it is appropriate  to use both ans\"atze to 
 study the systematic uncertainties in the fitting of the lattice QCD data.

As alluded to in Eq.~\refeq{eq:C_MB_MB_asymptotic}, the interpretation of the overlap of lattice states and interpolating fields acting on the vacuum state
involves the spinors $u_{B^*}$ and $u_{B}$ as follows
 \begin{align}
	\sum\limits_{\xvec}\,\frac{1}{4}(1+\gamma_0)\,\brackets{0\,|\,J^{\mu\alpha}_{B^*}(t,\xvec)\,\epow{i\Pvec'\xvec}\,|\,B^*,\Pvec,s=3/2,s_3} &=
	Z_{B^*}(\Pvec^2)\,V\,\delta_{\Pvec \Pvec'}\,u_{B^*}^{\mu\alpha}(\Pvec,s_3)
	\label{eq:lattice_overlap_Bstar}\\
%%%
	\sum\limits_{\yvec}\,\frac{1}{4}(1+\gamma_0)\,\brackets{0\,|\,J_{M}(t,\yvec)\,J^{\alpha}_{B}(t,\zvec)\,\epow{i\kvec'(\yvec-\zvec)}\,|\,M B,\kvec,s=1/2,s_3} &=
	Z_M(\kvec^2)\,Z_{B}(\kvec^2)\,V\,\delta_{\kvec \kvec'}\,u_{B}^{\alpha}(-\kvec,s_3)\,.
	\label{eq:lattice_overlap_MB}
\end{align}
We denote by $s$ the spin quantum number of the baryon fields, $s_{B^*} = 3/2$ and $s_{B} = 1/2$ (having $s_M = 0$ fixed for the pseudoscalar meson)
and by $s_3$ its projection to a specific axis.
The definition in Eq.~\refeq{eq:lattice_overlap_MB} assumes that the total linear momentum of the meson-baryon state is zero.

While the ratio $R^{B^*}_{M B}$ is constructed such that the numerical factors $Z_{B^*/M/B}$ cancel, the spinors remain in the fraction and are combined 
to spin sums $\Sigma^{B^*}_{M B}$ in the numerator and $\Sigma_{B^*}$ and $\Sigma_{B}$ in the denominator via summation over the third spin component.
To that end, we parametrize the slope $c_1 = ax$ in Eq.~\refeq{eq:fit_type_1}, \refeq{eq:fit_type_2} as follows
\begin{align}
	c_1 &= \sum\limits_{\sigma_3,\tau_3}\,\frac{\matelem(\Pvec=0, |\kvec|,\sigma_3,\tau_3)}{\sqrt{N_{B^*}\,N_{MB}}}\,V\,
	\frac{\tilde{\Sigma}^{B^*}_{M B}(\Pvec=0,\kvec,\sigma_3.\tau_3)}{\sqrt{\Sigma_{B^*}(\Pvec=0)\,\Sigma_{B}(|\kvec|)}}
	\label{eq:slope_matelem_parametrization}\\
	\tilde{\Sigma}^{B^*}_{M B}(\Pvec=0,|\kvec|,\sigma_3) &= \frac{1}{6}\,\sum\limits_{j=\pm 1,\pm 2,\pm 3}\,
	\mathrm{sign}(j)\,\frac{1}{4}(1+\gamma_0)_{\alpha\beta}\,u^{j \beta}_{B^*}(\Pvec=0,\sigma_3)\,\ubar^{\alpha}_{B}(|\kvec|\,\evec_j,\tau_3)\,,\nonumber
\end{align}
where we use the notation $\evec_{-j} = -\evec_j$ for $j = 1,2,3$ for brevity. The volume factor $V$ stems from the lattice Kronecker$-\delta$ in momentum space for the total linear
momentum $\Pvec$. We note that by our construction of the correlators, which are not summed over the source locations, but fulfill momentum conservation, we effectively
have to insert this factor by hand for correct normalization of the ratio. $N_{B^*}$ and $N_{M B}$ denote the normalization of the states and we use
the standard continuum-like normalization of on-shell states
\begin{align}
	\brackets{B^*,\Pvec,s_3\,|\,B^*,\Pvec',s_3'} &= \frac{E_{B^*}(\Pvec^2)}{m_{B^*}}\,V\,\delta_{\Pvec \Pvec'}\,\delta_{s_3 s_3'}\nonumber\\
	\brackets{M B,\kvec_M,\kvec_B,s_3\,|\,M B,\kvec_M',\kvec_B',s_3'} &= 2\,E_{M}(\kvec_M^2)\,V\,\delta_{\kvec_M \kvec_M'}\times
	\frac{E_B(\kvec_B^2)}{m_B}\,V\,\delta_{\kvec_B \kvec_B'}\,\delta_{s_3 s_3'}\nonumber\\
	N_{B^*} &= V\,,\quad N_{M B} = 2\,V^2 \frac{E_M(\kvec^2)\,E_B(\kvec^2)}{m_B}\,.
	\label{eq:normalization_of_states}
\end{align}
In the last line of equation \refeq{eq:normalization_of_states} we specialized to the case at hand with $\Pvec=0$ and $\kvec_M = -\kvec_B = \kvec$.\\

To sum the spinors in the numerator we likewise parametrize the matrix element according to leading order effective field theory,
\begin{align}
	\matelem(\Pvec_{B^*},\kvec_{M},\kvec_{B},\sigma_3,\tau_3) &=
	C_\mathrm{CG}\,\frac{g^{B^*}_{M B}}{2 m_B}\,\ubar^{\mu\alpha}_{B^*}(\Pvec_{B^*},\sigma_3)\,k_{M \mu}\,u_B^\alpha(\kvec_B,\tau_3)\,.
	\label{eq:matelem_parametrization}
\end{align}
$C_{\mathrm{CG}} = C_\mathrm{CG}(I_{B^*},I_{B^*}^3\,|\,0,0; I_B,I_B^3)$ is the Clebsch-Gordan coefficient for the coupling of the isospin of $M$ and $B$ to match that of $B^*$.\\

With Eq.~\refeq{eq:matelem_parametrization} we can then use the standard spin sums for spin-1/2 fermions and the Rarita-Schwinger 
field\cite{Nozawa:1990gt,Alexandrou:2007dt}
\begin{align}
	\Sigma_{B} (|\kvec|) &= \frac{E_B(\kvec^2) + m_B}{2 m_B}\nonumber\\
	\Sigma_{B^*} (\Pvec=0) &= \frac{2}{3} \nonumber\\
	\Sigma^{B^*}_{M B}(\Pvec=0,|\kvec|) &= \Sigma_{B} (|\kvec|) \times \Sigma_{B^*} (\Pvec=0) =
	\frac{1}{3}\,\frac{E_B(\kvec^2) + m_B}{m_B}\,.
	\label{eq:spin_sums}
\end{align}

\noindent
We thus write the coupling as
\begin{align}
	g^{B^*}_{M B} &= c_1 \,\frac{\sqrt{N_{B^*}\,N_{M B}}}{V C_\mathrm{CG}}\,\frac{2 m_B}{|\kvec|}\,\left( \frac{1}{3}\,\frac{E_B(\kvec^2) + m_B}{m_B} \right)^{-1/2}\,.
	\label{eq:coupling_formula}
\end{align}
With this expression we can go back and rewrite the matrix element in terms of the extracted slope $c_1$,
\begin{align}
	|\matelem(\Pvec=0,|\kvec|^2)|^2 &= C_{\mathrm{CG}}^2\,\left( \frac{g^{B^*}_{M B}}{2m_B} \right)^2\,\frac{2}{3}\,\kvec^2\,\frac{E_B(\kvec^2) + m_B}{m_B} 
	= c_1^2\,\frac{2 N_{B^*} N_{M B}}{V^2}\,.
	\label{eq:matelem_formula}
\end{align}
We note, that the expression in Eq.~\refeq{eq:matelem_formula} gives the squared matrix element for the transition between a certain isospin state of $B^*$ and
a certain product of isospin states for $M$ and $B$, such that
\begin{align}
C_\mathrm{CG} &= \brackets{ I_{B^*},\,I_{B^*}^3\,|\, I_{M},\,I_{M}^3,\, I_{B},\,I_{B}^3}\,.
\label{eq:c_clebsch_gordan}
\end{align}
In Table~\ref{tab:isospin} we give the isospin values for the decuplet resonances and their decay channel.

\begin{table}
\begin{center}
\begin{tabular}{cc|cc|cc|c}
$B^*$         & $I_{B^*},\,I_{B^*}^3$ & $M$      & $I_{M},\,I_{M}^3$ & $B$        & $I_{B},\,I_{B}^3$ & $|C_\mathrm{CG}|$ \\
\hline
$\Delta^{++}$ & $3/2,\,+3/2$          & $\pi^+ $ & $1,\,+1$          & $N^+$      & $1/2,\,+1/2$      & $1$               \\
$\Sigma^{*+}$ & $1,\,+1$              & $\pi^+ $ & $1,\,+1$          & $\Lambda$  & $0,\,0$           & $1$               \\
$\Sigma^{*+}$ & $1,\,+1$              & $\pi^+ $ & $1,\,+1$          & $\Sigma^0$ & $1,\,0$           & $\sqrt{1/2}$      \\
$\Xi^{*-}$    & $1/2,\,-1/2$          & $\pi^- $ & $1,\,-1$          & $\Xi^0$    & $1/2,\,+1/2$      & $\sqrt{2/3}$      \\
\hline
\hline
\end{tabular}
\end{center}
\caption{We give the isospin quantum numbers of the decuplet (first column) and the  two-particle decay channel consisting of a   meson with quantum number given  in the second column and a spin-1/2 baryon with quantum numbers given in the third column. In the last column we give the absolute value of the isospin factor $C_{CG}$. }
\label{tab:isospin}
\end{table}

\subsection{Density of states} To apply Fermi's Golden Rule we need to estimate the density of states at the transition energy.  For a
free pion (pseudoscalar $M$) and a free baryon $B$ in the center-of-mass frame with $k = |\kvec|$ the  total energy is
\begin{align*}
 E_f(k) = E_M(k) + E_B(k) = \sqrt{m_M + k^2} +  \sqrt{m_B + k^2}.
\end{align*}
Furthermore, we assume an isotropic density in the volume $L^3$, with the unit cell in momentum space being of size $2\pi/L$.
Up to momentum $k$ we thus count $\Omega(k) / (2\pi/L)^3 = 4\pi k^3/3 / (2\pi/L)^3$ states. Varying $k$ we then have a density of states given by
\begin{align}
\rho(E_f) &= \frac{d\Omega}{dE_f} \,\frac{L^3}{8\pi^3} = \frac{L^3}{2\pi^2}\,k^2\,\frac{dk}{dE_f} = \frac{L^3}{8\pi^2}\,k\,\frac{E_f^4 - (m_M^2-m_B^2)^2}{E_f^3}\,.
\label{eq:density_of_states}
\end{align}

\subsection{Decay width to leading order} Having the overlap $x$ from the lattice correlator functions and using the density of states we can  connect, to leading order in the effective theory,  the decay width in the continuum to that  on the lattice by suitable normalization. To that end we
observe, that to leading order in the continuum effective field theory, we have
\begin{align}
\Gamma &= \frac{1}{2 s_{B^*}+1} \, \frac{m_{B^*}}{E_{B^*}}\,\sum\limits_{s_{B^*},s_B}\,|\matelem(s_{B^*}, s_B)|^2\,\int\,d E_f\,\frac{m_B}{E_B\,2 E_M}\,
\left[ \frac{k(E_f)^2}{2\pi^2}\,\frac{d k(E_f)}{d E_f}\right]\,2\pi\,\delta(E_f - E_i)\,.
\label{eq:width_lo_eft}
\end{align}
Evaluating the $\delta$-functional in the center-of-mass frame with $E_i = m_{B^*} = E_{B^*}$ we obtain
\begin{align}
\Gamma &= 2\pi\,\left[ \frac{1}{2 s_{B^*}+1}\,\sum\limits_{s_{B^*},s_B}\,|\matelem(s_{B^*}, s_B)|^2\,\right]\,\frac{V^3}{N_{MB}\,N_{B^*}}\,\frac{1}{V}\,\rho(E_i) \nonumber\\
&= 2\pi\,\left[ \frac{2\,c_1^2}{2 s_{B^*}+1}\right]\,\rho(E_i) \nonumber\\
&= 2\pi\,\brackets{|\matelem|^2}\,\rho(E_i)\,.
\label{eq:width_lattice_formula}
\end{align}
We note that the expression of Eq.~\refeq{eq:width_lattice_formula} contains the sum over all final states (in particular all spin configurations of the field $B$)
and the average over all initial spin states of the spin-3/2 baryon $B^*$.
In Eq.~\refeq{eq:width_lattice_formula},  the width calculated  is independent of the normalization of states chosen at intermediate stages, as one would expect.
The coupling in Eq.~\refeq{eq:coupling_formula}, on the other hand, carries an explicit dependence on the normalization of states shown by the appearance of the factor
$\sqrt{ N_{B^*}\,N_{M B} }$.

%%% {\bf Dina: do we really want to make this a subsection?}
%\subsection{Excited states}

We would like to note that in a realistic setup the lattice Hilbert space is of high dimension and the lattice transfer matrix correspondingly large. The restriction to
a two-dimensional subspace may still be justified, if the first excited 
states in the $B^*$ and $M\,B$ channels are  sufficiently higher in energy.
As usual this would lead to an exponential suppression of contributions from such states
as assumed in the ellipsis in Eq.~\refeq{eq:derivation_fit_formula}.
%%% {\bf Dina: please check because I may have  not understood your point completely}
Only in this case 
the overlap is proportional to  the time separation, receiving
contributions from single transitions from initial to final state anywhere along the time axis.

%%%%%%%%%%%%%%%%%%%%%%%%%%%%%%%%%%%%%%%%%%%%%%%%%%%%%%%%%%%%%%%%%%%%%%%%%%%%%%%%%%%
% Numerical results
%%%%%%%%%%%%%%%%%%%%%%%%%%%%%%%%%%%%%%%%%%%%%%%%%%%%%%%%%%%%%%%%%%%%%%%%%%%%%%%%%%%

%%% \input{numericalResults}

\section{Numerical results}
\label{sec:Numerical results}
We analyze two ensembles:
%one for a hybrid action with domain wall valence quarks on a $N_f=2+1$ staggered sea \cite{Bernard:2001av}
%and one with $N_f=2+1$ domain wall fermions with the
%unitary action~\cite{Arthur:2012opa}. 
%%%%%%%%%%%%%%%%%%%%%%JWN
one for a hybrid action with domain wall valence quarks on a $N_f=2+1$ staggered sea~\cite{Bernard:2001av} and  $m_\pi =350\mev$
and one for a unitary action with $N_f=2+1$ domain wall quarks~\cite{Arthur:2012opa} and
 $m_\pi =180\mev$.   
Subsequently we will use the labels ``hybrid'' and ``unitary'' to distinguish results obtained using these two sets of gauge configurations. Results for the $\Delta$ resonance for the hybrid calculation have been  reported in Refs.~\cite{Alexandrou:2013ata,Alexandrou:2014qka} and thus  we do not discuss them in detail
here.\\

%%%%%%%%%%%%%%%%%%%%%%%%%%%%%%%%%%%%%%%%%%%%%%%%%%%%%%%%%%%%%%%%%%%%%%%%%%%%%%%%%%%
%   Simulation details
%%%%%%%%%%%%%%%%%%%%%%%%%%%%%%%%%%%%%%%%%%%%%%%%%%%%%%%%%%%%%%%%%%%%%%%%%%%%%%%%%%%
\subsection{Simulation details}

For the hybrid setup  we use an ensemble of  $N_f=2+1$ staggered fermion configurations
with the light quark mass corresponding to a pion mass of $350\mev$ and the strange quark mass fixed to its physical value. This MILC ensemble
is labeled as  \textit{MILC\_2864\_m010m050} \cite{Bazavov:2009bb}. As
valence quarks we consider
domain wall fermions with the light bare quark mass adjusted to reproduce
the lightest pion mass obtained using $N_f=2+1$
staggered quarks~\cite{Edwards:2005ym}.
The valence strange-quark mass was set using the $N_f=3$ ensemble
by requiring the valence pseudoscalar mass to be equal to the mass
of the Goldstone boson constructed using
 staggered quarks~\cite{WalkerLoud:2008bp,Alexandrou:2010jv}.
% Technical details
%of this tuning procedure 
%are given in Refs.~\cite{WalkerLoud:2008bp,Renner:2004ck,Hagler:2007xi}.
For the unitary setup we use an ensemble of gauge configurations generated by the RBC-UKQCD collaborations with $N_f = 2 + 1$ domain-wall fermions
and the Iwasaki gauge-action  %employed. %%%%%%%%%%JWN
 labeled as \textit{RBC\_b1p75\_L32T64\_m045m001}~\cite{Arthur:2012opa}.
The simulation parameters for both cases are given in Table \reftab{tab:1}.
\begin{table}
\centering
  \begin{tabular}{l|cccc|cc|cc}
  \hline
  action & $ L^3\times T$ & $\mps / \mev$ & $a / \fermi$ & $L_5$ & $N_\mathrm{conf}$ & $N_\mathrm{src}$ & $\alpha_\mathrm{APE} / N_\mathrm{APE}$ & $\kappa_\mathrm{Gaussian} / N_\mathrm{Gaussian}$ \\
  \hline
  hybrid  & $28^3\times 64$ & $350$ & $0.124$ & 16 & 210 & 4 indep. & $2.0 / 20 $ & $4.0    / 50$ \\
  unitary & $32^3\times 64$ & $180$ & $0.143$  & 32 & 254 & 4 coh.   & $2.5 / 25 $ & $0.5625 / 70$ \\
  \hline
\end{tabular}
%\caption{
%  In the second, third, fourth and fifth columns we give the lattice size, the pion mass, the lattice spacing
%  and the size of the fifth-dimension for the two ensembles considered in this work.
%  In the sixth and seventh columns we give the number of configurations used for the analysis
%  and the number of source positions per configuration.
%  In the last two  columns we give the parameters for the APE  and the Gaussian smearing used in the construction of the interpolating fields.
%}
%%%%%%%%%%%%%%%%%JWN
\caption{Simulation parameters.
  The second, third, fourth and fifth columns give the lattice size,  pion mass, lattice spacing
  and size of the fifth-dimension for the two ensembles considered in this work.
  The sixth and seventh columns give the number of configurations used for the analysis
  and the number of source positions per configuration.
  The last two  columns give the parameters for the APE  and the Gaussian smearing used in the construction of the interpolating fields.
}
\label{tab:1}
\end{table}
For the hybrid ensemble we perform four independent measurements on 210 gauge configurations. The source locations for these measurements are separated by $T/4$ in time direction
and the spatial coordinates are randomly chosen across the spatial volume. In the case of the unitary ensemble we use 4 independent propagators, which are inserted
coherently into a single sequential source. Upon subsequent inversion of the Dirac operator the latter gives rise to a superposition of four sequential propagators at distance
$T/4$ in time direction and thus four coherent sets of contractions.

We use source- and sink-smearing on all interpolating fields to improve the overlap of our interpolating fields with the ground state. The forward  and sequential
propagators are smeared using
Gaussian smearing with the APE smeared gauge links entering in the hopping matrix of the
Gaussian smearing function. The smearing parameters for both lattices are given in Table~\reftab{tab:1}.

Inversions of the Dirac operator have been performed using the packages \textit{QUDA}  \cite{Clark:2009wm,Babich:2011np}
for the hybrid calculation
and \textit{Qlua}~\cite{qlua:2015}  using Moebius-accelerated domain wall fermions for the unitary action \cite{Yin:2011np}.

\begin{figure}[htb]
\centering
\includegraphics[width=0.8\textwidth]{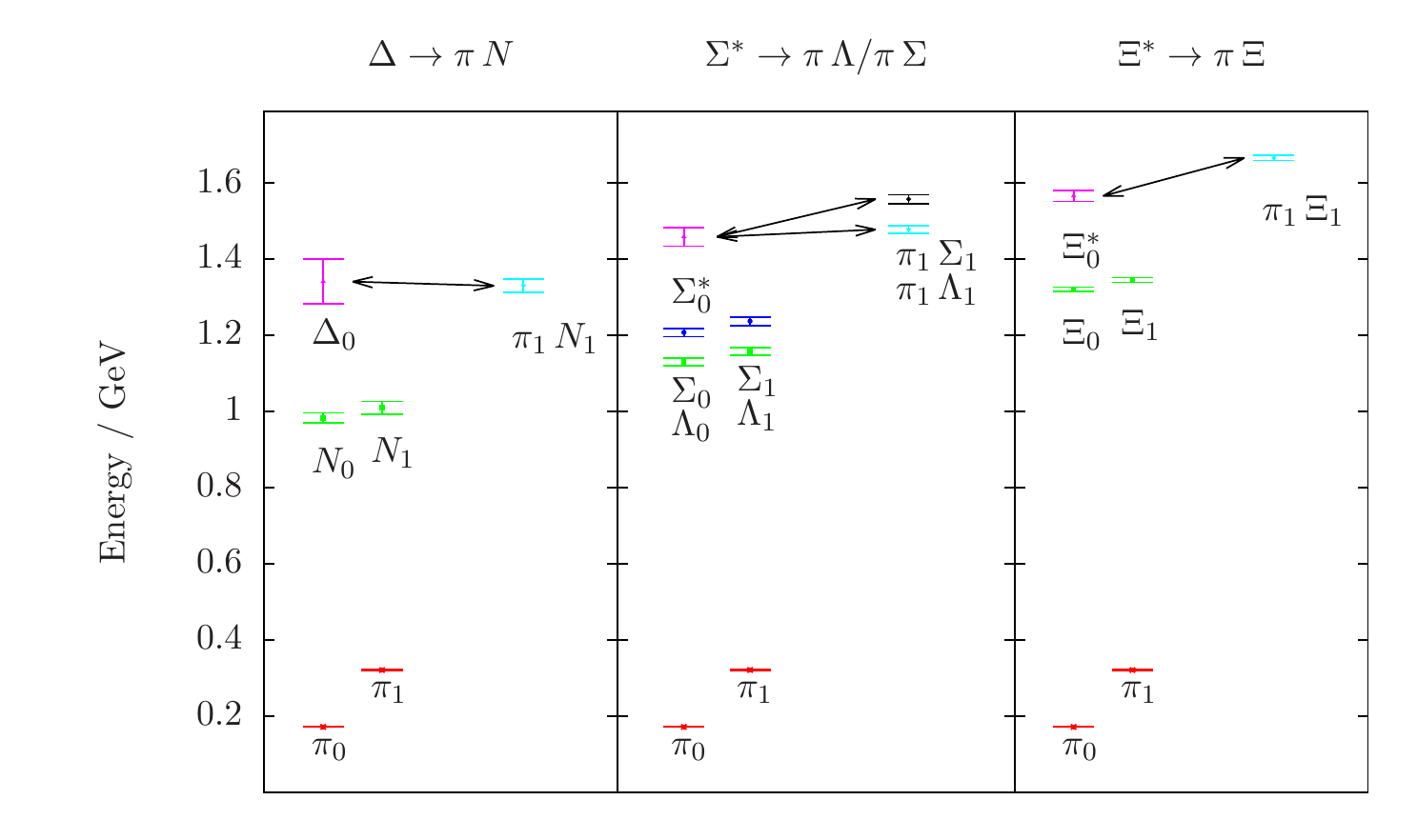}
\caption{
% Energies of the relevant states entering in the study of the decays for the unitary ensemble.
%%%%%%%%%%%%%%%%%%%%%JWN          Fig.~\reffig{fig:3a}   fig3
  Energies of the states entering the study of the decays for the $m_\pi$ = 180 MeV unitary ensemble.
  The black  arrows mark the transitions we consider in this work (cf. Eq.~\refeq{eq:decuplet_decays}).
}
\label{fig:3a}
\end{figure}

In Fig.~\reffig{fig:3a} we show  the energies of the states
that are relevant for our calculation. The energies  for zero and one unit of momentum
$|\qvec| = 2\pi/L$ are shown. We use a notation analogous to reference \cite{McNeile:2002fh}  giving
the one-particle interpolating fields a subscript labeling their momentum, i.e. $\pi_0$ denotes
the pion-state with zero momentum, $\pi_1$ with one unit of momentum etc. Likewise
$\pi_1\,N_1$ is the pion-nucleon state, where each interpolating field is constructed with one unit
of momentum, while keeping zero total momentum.
With the label $\pi \, B$, where $B = N,\,\Lambda,\,\Sigma,\,\Xi$ we
denote the sum of the individual energies of
the pion and the octet baryon $B$. 
The individual energies are determined from the two-point correlators of each particle.

\begin{figure}[htb]
\centering
\includegraphics[width=0.8\textwidth]{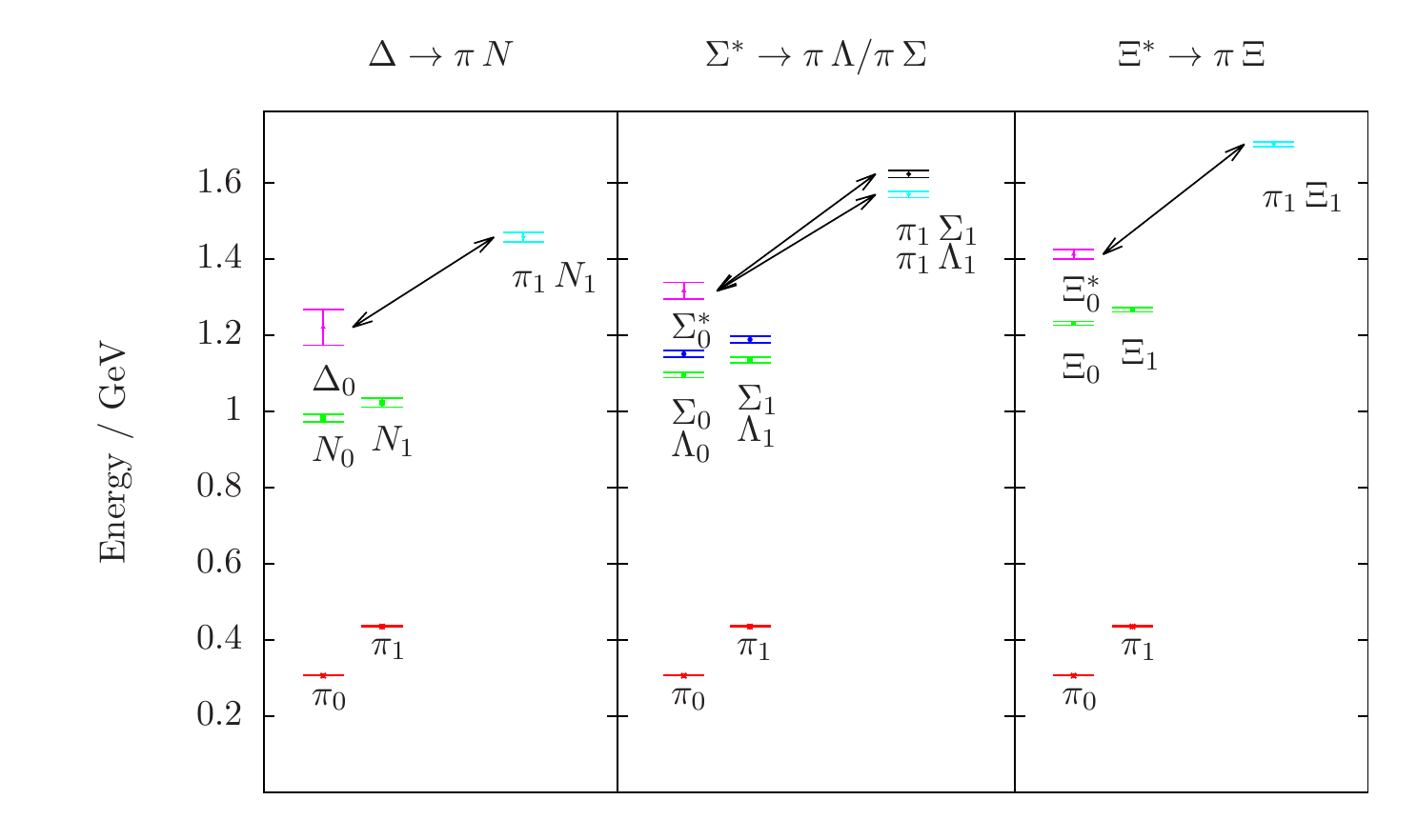}
\caption{
%  Energies of the relevant states entering in the study of the decays for the hybrid ensemble.
  %%%%%%%%%%%%%%%%%%%%%JWN	 Fig.~\reffig{fig:3b}   fig4
  Energies of the states entering the study of the decays for the $m_\pi = 350 \mev$ hybrid ensemble.
  The black arrows mark the transitions we consider in this work (cf. Eq.~\refeq{eq:decuplet_decays}).
} 
\label{fig:3b}
\end{figure}
%%%%%%%%%%%%%%%%%%%%%%%%JWN start edit
For the unitary ensemble we observe a near degeneracy of energy levels for the $\Delta$ and the $\pi\,N$ scattering state
as well as for the $\Sigma^{*}$ and the  $\pi\,\Lambda$  scattering state.
On the other hand, a significant energy gap
exists between the  $\Sigma^*$  and the $\pi\,\Sigma$  scattering state, and the $\Xi^*$ and  the 
$\pi\,\Xi$  scattering state.
The situation is qualitatively different for the hybrid calculation, 
for which the relevant spectrum is shown in Fig.~\reffig{fig:3b}.
We observe a larger energy gap for all transitions under consideration, which is roughly the same in all cases.

%%%%%%%%%%%%%%%%%%%%%%%%JWN new:
This qualitative difference arises from the larger values of $m_\pi$.  In our approximation $ E_{MB} = E_M + E_B$,
so the gap is $\delta = E_{B^*} - E_{MB}  =  E_{B^*} - E_M - E_B $.
Considering Fig.~\reffig{fig:3b} for the case  $\Delta  - \pi N$ one observes
that at  $q = 0$, $E_N +E_\pi $ is significantly greater than $E_\Delta$ so $\delta$ is negative and far from threshold.
Taking into account that $E_N +E_\pi$  increases with $q$, and so does $|\delta|$,
the gap gets even bigger as compared to $|\delta|$ at zero $q$.
The same thing happens for $\Sigma^*$ and for $\Xi^*$. 
In contrast, in Fig.~\reffig{fig:3a} for the ensemble at $180 \mev$ pion mass, we see that the $\Delta$ is unstable
and there is a chance $\delta$ will pass through zero at the relevant $q$ for the transition. Indeed we see that $\delta$ is small and slightly positive.
Similarly, for $\Sigma^*$, $\delta$ is small and slightly negative.
%%%%%%%%%%%%%%%%%%%%%%%%JWN end edit

%%%%%%%%%%%%%%%%%%%%%%%%%%%%%%%%%%%%%%%%%%%%%%%%%%%%%%%%%%%%%%%%%%%%%%%%%%%%%%%%%%%
%   Delta to pi N
%%%%%%%%%%%%%%%%%%%%%%%%%%%%%%%%%%%%%%%%%%%%%%%%%%%%%%%%%%%%%%%%%%%%%%%%%%%%%%%%%%%
\subsection{$\Delta \to \pi N$}
\label{subsec:delta_to_pi_n}
\begin{figure}[htpb]
\begin{center}
\includegraphics[width=0.7\textwidth]{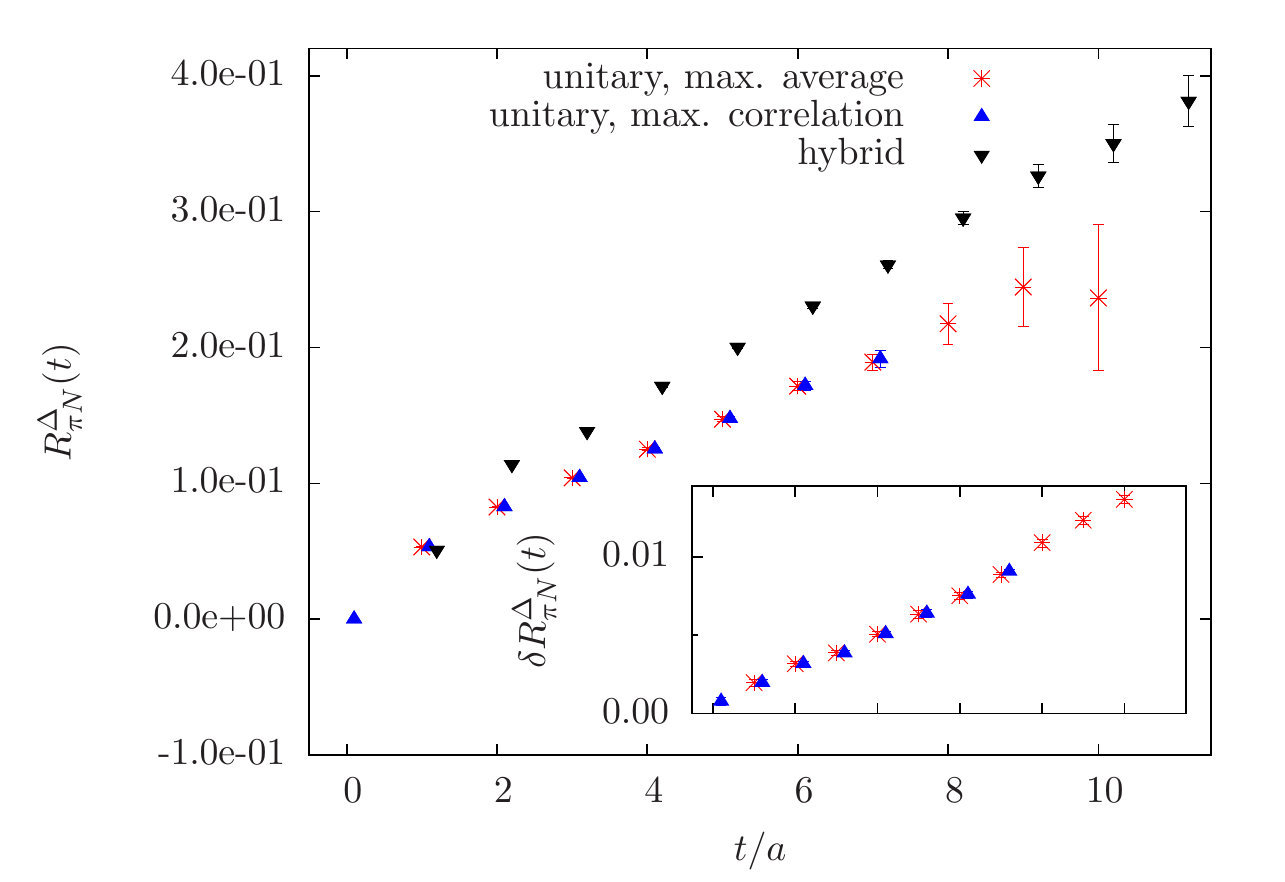}
\end{center}
\caption{
  Ratio $R^{\Delta}_{\pi N}$ for the unitary and hybrid calculation;
  the detail plot shows  the evolution of the statistical uncertainty with $t/a$. The errors
  are computed using the $\Gamma$-method~\cite{Wolff:2003sm}. The red crosses
  are obtained using maximal averaging to construct the ratio for the unitary action,
  while the blue triangles exploit maximal correlation.
  The black triangles show the ratio obtained with the
  hybrid action using maximal correlation combination of data. Data points for different curves have been
  displaced horizontally.
}
\label{fig:4}
\end{figure}

We first discuss the case of the $\Delta$ resonance.
As shown in Fig.~\reffig{fig:3a}, the lattice kinematics produce a scattering $\pi N$ state
that is approximately degenerate with the $\Delta$ mass, thus satisfying
one of the conditions for the validity of the method. For the hybrid action,
used in our study, the energies have a sizeable gap as shown in Fig.~\reffig{fig:3b}.
In  Fig.~\reffig{fig:4} we show the ratio $R^{\Delta}_{\pi\,N}$ for both the unitary and the hybrid action.

For the unitary case we compare two different ways to combine the available lattice data. In the first one, all individual factors
of the two-point correlators entering the ratio are averaged before the ratio is built
using the maximal set of lattice symmetries for the individual correlators.
This way of combining data will benefit from possible cancellation of additive lattice artifacts in
individual correlation functions.  We refer to it as maximal averaging.
In the  second approach, we build the ratio for each data set given by the tuple
(momentum / direction, forward and backward propagators, source location)
and in the final step combine the individual estimates for the ratio. Since the correlator data within one and the same tuple
is maximally correlated, such an average would benefit error cancellations due to statistical correlation.

We note, that due to the coherent source method used with the unitary action,
we must keep the source-sink time separation sufficiently smaller than the distance between the source insertions, i.e.
$t_f-t_i \ll T/4$. The ratio $R^\Delta_{\pi N}$ shown in  Fig.~\reffig{fig:4} exhibits a time dependence
that is consistent with the expected linear behavior
for both hybrid and unitary action, as well as for both types of averages.
An overall comparison of the two approaches used for constructing the ratio with
the unitary action  does not reveal any significant
difference in the mean value or the statistical uncertainty of the data points where they are both defined.
However, on closer examination, the maximally averaged approach
produces data
for larger time slices. This is due to the fact that the $\Delta$ and nucleon correlators are more accurately determined
having thus a lower probability of becoming non-positive in the sampling part of
the error estimate. We shall thus use  maximal averaging  to combine data in what follows.

\begin{figure}[htpb]
\begin{center}
\includegraphics[width=0.8\textwidth]{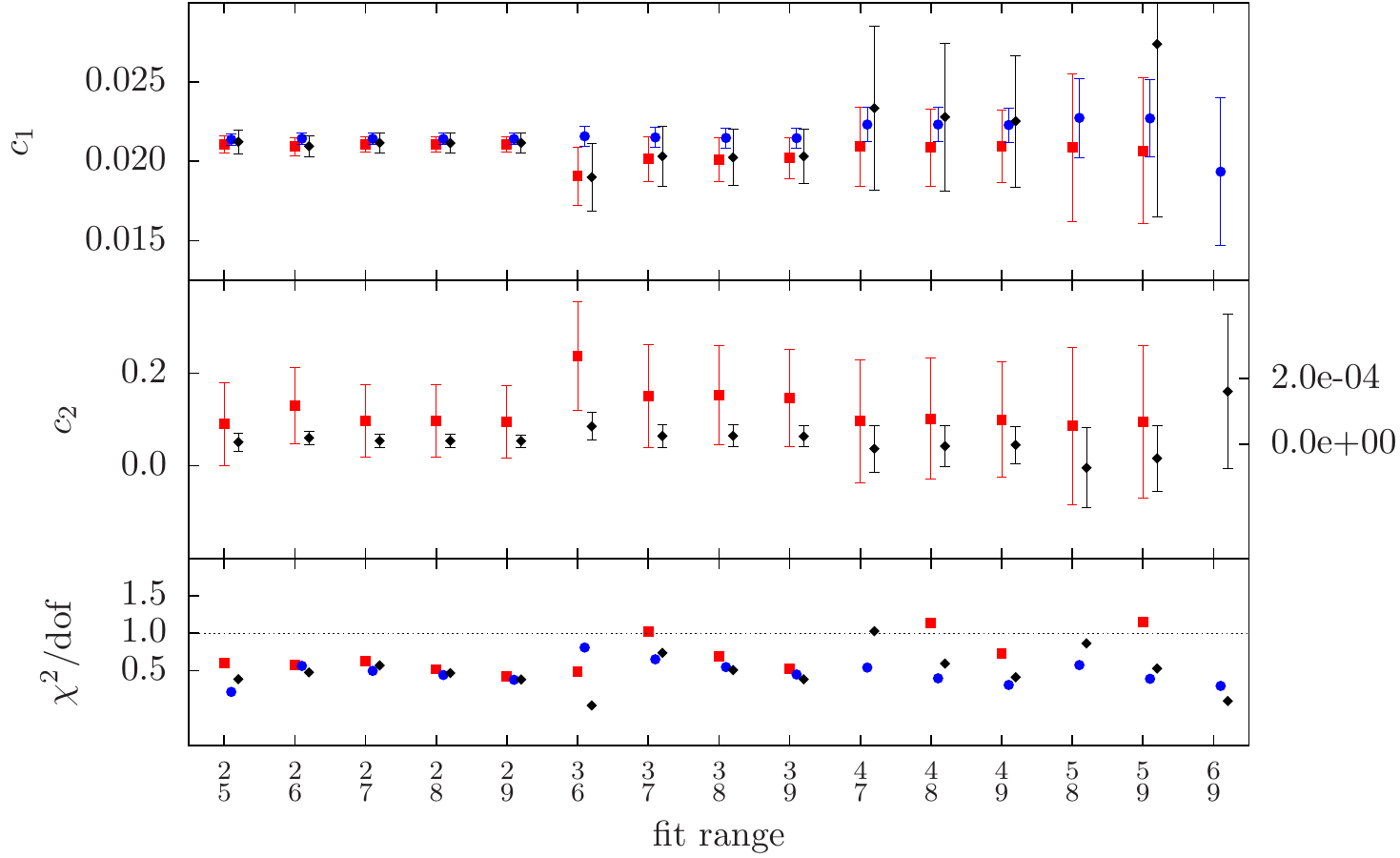}
\end{center}
\caption{
  We show the values of the parameters $c_1$ (upper panel), $c_2$ (central panel) and $\chisqrPerDof$ for the fit (lower panel)
  using  three different ans\"atze  to fit  $R^\Delta_{\pi\,N}$. The symbols  represent from left to right:
  type 1 fits using $f_1(t)$  with 3 fit parameters (red squares),
  type 2 using $f_2(t)$ with 2 fit parameters (blue circles) 
  and type 3 using $f_2$ with 3 fit parameters (black diamonds).
  The x-label gives the fit interval as a column $t_\mathrm{min}$ atop $t_\mathrm{max}$
  in lattice units;
  in the case of $c_2$ we show results using type 1 with the labels to the left $y$-axis and type 2 with respect to right $y$-axis.
}
\label{fig:5} 
\end{figure}

In Fig.~\reffig{fig:5} we show the results from fitting  $R^{\Delta}_{\pi\,N}$ using the two fit ans\"atze given in Eqs.~\refeq{eq:fit_type_1},\refeq{eq:fit_type_2},
for the unitary action.
We observe that the linear fit ansatz labeled as type 2, which uses a correlated fit with the function  $f_2(t)$ and two free parameters $c_0,\,c_1$,
already leads to fits with a value for $\chisqrPerDof$ below one (bottom panel).
The fit value for the slope determined by $c_1$  does not show significant variation
when scanning the fit ranges  from $\left[ 2,\,5\right]$ to $\left[ 6,\,9 \right]$.
As shown  in Fig.~\reffig{fig:4}, the statistical uncertainty of the fit parameters increases with increasing the lower fit
range to larger values as expected from
the dependence of the statistical uncertainty on the fitted data. Moreover, using as a fit function $f_1(t)$ 
we do not observe a significant change for $c_1$. Neither do we observe any significant dependence of the central value
of $c_1$ on the number of parameters. 
In fact, $c_2$ is
statistically consistent with zero for both ans\"atze $f_1$ and $f_2$  as shown in the center panel of Fig.~\reffig{fig:5}.
We point out  that we include different labels on  the left and right $y$-axes
to show  the values extracted using the  fitting functions $f_1$ and $f_2$, respectively,
because the order of magnitude of $c_2$ differs.

As indicated in Fig.~\reffig{fig:5}, we perform a large number of fits with different fit types
using the fit ans\"atze $f_{1,2}(t)$ and different time ranges for
the estimate of the slope and the energies, for zero and one unit of momentum that enter
the calculation. We note that with the $\Delta$ at rest, its mass  does not enter the calculation of the coupling constant
%nor   JWN
or
the width. This result relies on energy conservation and generalizes to
all decays studied here.

Since many fits %show JWN
yield an acceptable value for $\chi^2/\mathrm{dof}$, we combine the different analyses with an appropriate weight
to extract a mean value and meaningful estimates for the statistical and systematic uncertainty to account for the varying
goodness of the fits and precision of the estimates from them.
We consider the distribution of the results from each individual fit and associate a weight to it as follows:
\begin{align}
	w(g,\,\Gamma) &= \prod\limits_{A} \,w(A) \nonumber\\
	w(A) &= \left( 1 - 2\,|0.5 - p_A| \right)^2 \times \mathrm{var}({A})^{-1}\,.
	\label{eq:weight}
\end{align}
Here,   %%%%% added
$p_A$ denotes the p-value for the fit of quantity $A$,
\begin{align*}
  p_A &= \int\limits_{\chi^2_A}^{\infty}\,f_{\chi^2,\,\mathrm{dof}}(X)\,dX\,,
\end{align*}
where $\chi^2_A$ is the observed value of $\chi^2$ for the fit of quantity $A$ and $f_{\chi^2,\,\mathrm{dof}}$ the density
function for the $\chi^2$ distribution for dof degrees of freedom.
$A$ runs over all the quantities %, 
that have been derived from a fit and enter the calculation of $g$ (or $\Gamma$), i.e. 
the slope parameter $c_1$, the meson mass $m_M$ and baryon masses $m_B$. The definition in Eq.~\refeq{eq:weight} gives a higher weight
to fits with a p-value close to 0.5, such that there is equal probability of finding results above and below the observed fit value, and
to those fits with smaller variance of the fit result.
We then
take a weighted average from the distribution as the mean value,
\begin{align*} 
  \bar{g} &=  \sum\limits_{i} \,g_i \,w(g_i) \times \left( \sum\limits_i\,w(g_i) \right)^{-1}\,,
\end{align*}
where the sum runs over all fits labeled by index $i$.
The statistical uncertainty is calculated from the variance of the bootstrap samples for the weighted mean,
\begin{align*}
  \delta\,g_\mathrm{stat} &= \sqrt{\mathrm{var}\left(\bar{g}\right)}\,.
\end{align*}

Finally, the systematic uncertainty is estimated from the variance of the weighted distribution of the set $\{g_i\}$: we form a histogram,
where each $g_i$ gives a count proportional to $w(g_i)$ to the corresponding bin. The square root of the variance derived from this distribution
gives the systematic error $\delta\,g_\mathrm{sys}$.
We then quote our results as
\begin{align*}
  g &= \bar{g} \,\left(\delta\,g_{\mathrm{stat}}\right)\,\left( \delta\,g_{\mathrm{sys}} \right)\,.
\end{align*}
We proceed in the same way for the evaluation of the width  $\Gamma$. 
Following this procedure, we arrive at the values given in Eqs. \refeq{eq:g_deltapp2piN_unitary} and \refeq{eq:width_deltapp2piN_unitary}.
\begin{align}
	g^{\Delta}_{\pi N}\left(\mathrm{unitary} \right) &= 23.7 \,( 0.7 )\,(1.1 )
\label{eq:g_deltapp2piN_unitary}\\
a\Gamma^{\Delta}_{\pi N}\left(\mathrm{unitary} \right) &= 0.0868 \,(57)\,( 32 )
\label{eq:width_deltapp2piN_unitary}
\end{align}

%%%%%%%%%%%%%%%%%%%%%%%%%%%%%%%%%%%%%%%%%%%%%%%%%%%%%%%%%%%%%%%%%%%%%%%%%%%%%%%%%%%
%   Sigma^* to pi Lambda
%%%%%%%%%%%%%%%%%%%%%%%%%%%%%%%%%%%%%%%%%%%%%%%%%%%%%%%%%%%%%%%%%%%%%%%%%%%%%%%%%%%
\subsection{$\Sigma^* \to \pi \Lambda$}

\begin{figure}[htpb]
\begin{center}
\includegraphics[width=0.7\textwidth]{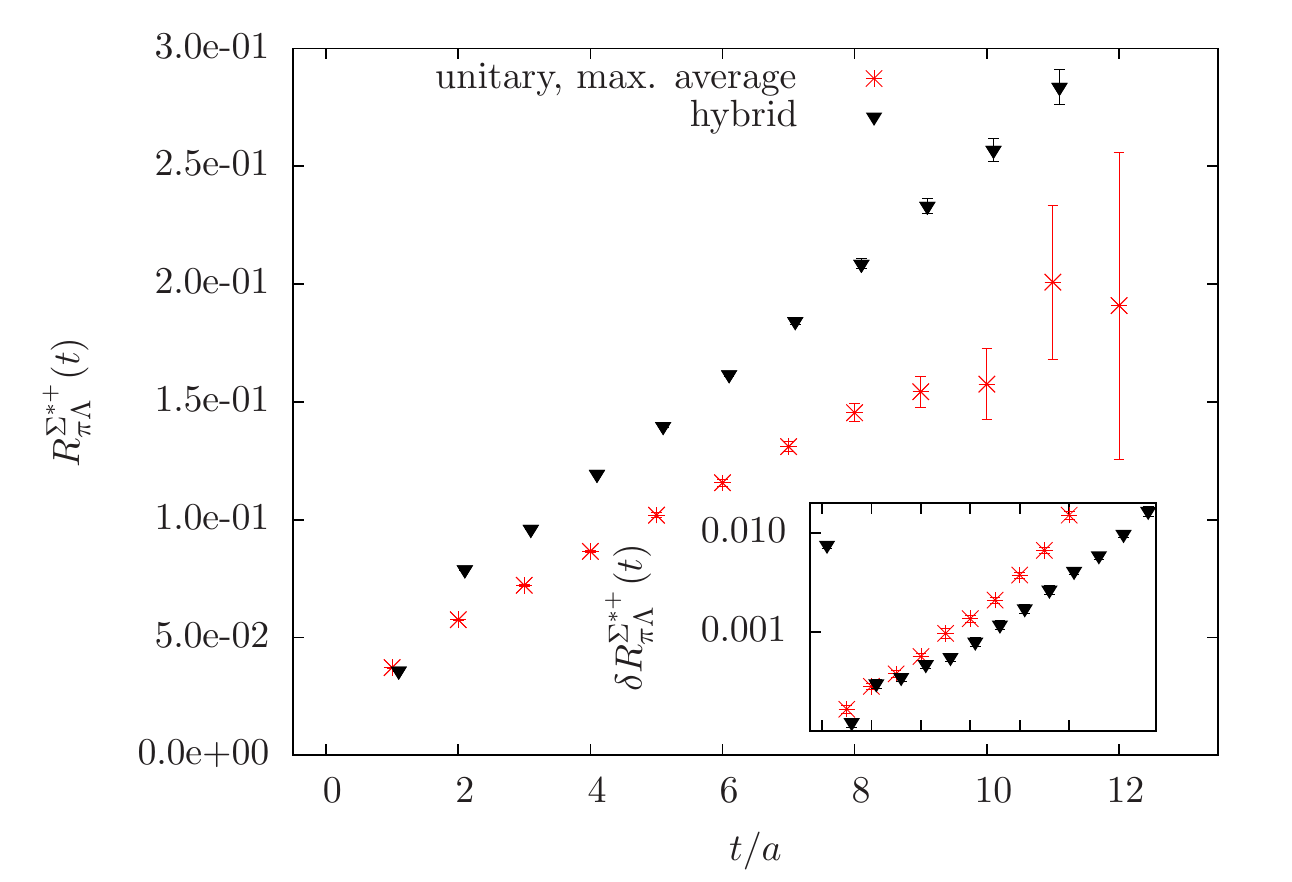}
\end{center}
\caption{
Ratio $R^{\Sigma^{*+}}_{\pi^+ \Lambda}$ for the unitary and hybrid calculations.  %%% s added
The notation is the same as that in Fig.~\ref{fig:4}.
}
\label{fig:6}
\end{figure}

\begin{figure}[htpb]
\begin{center}
	{\normalsize
          \includegraphics[width=0.8\textwidth]{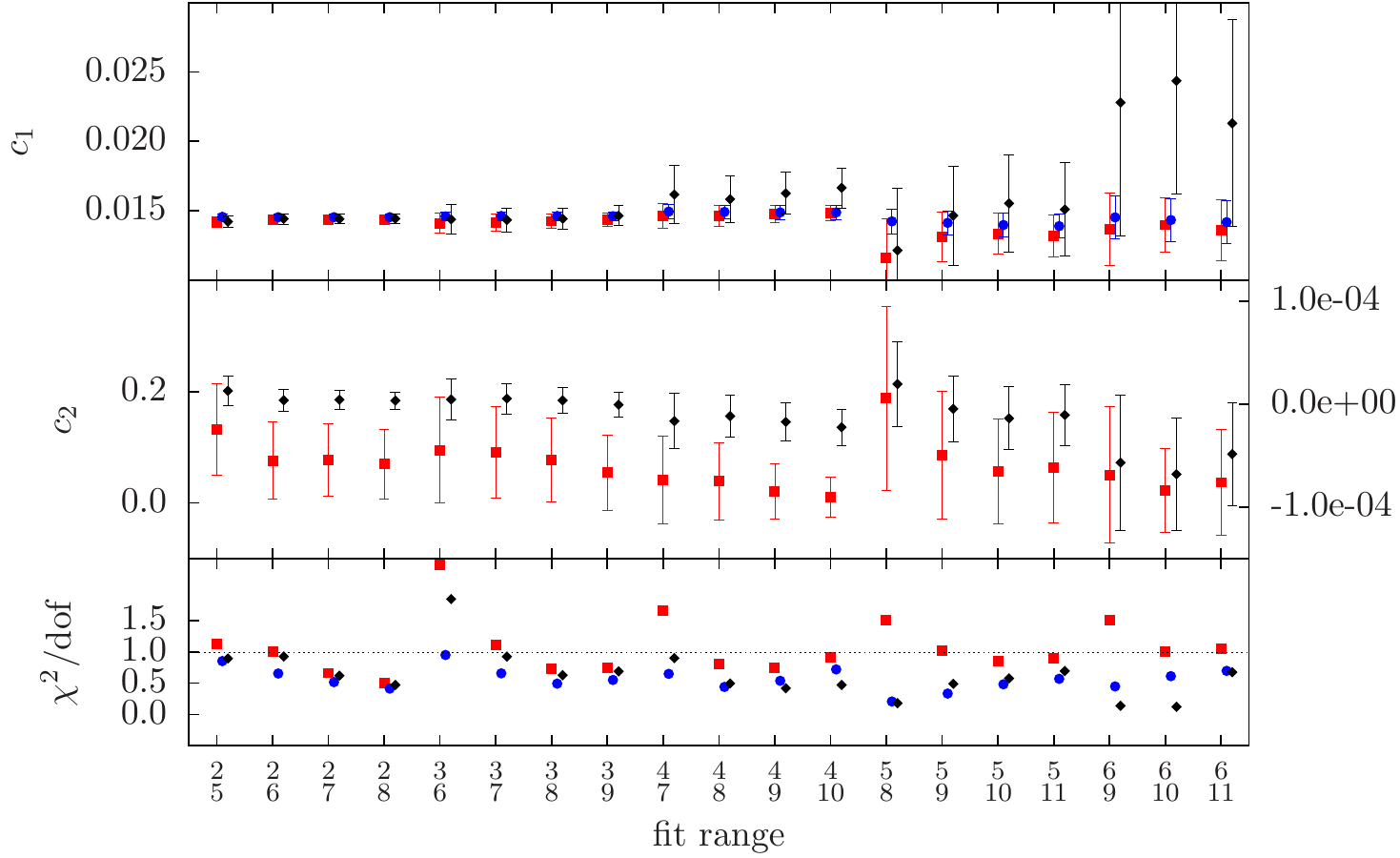}
}
\end{center}
\caption{
  Fit range dependence of the fit parameters $c_1,\,c_2$ for three different fits
  of ratio $R^{\Sigma^{*+}}_{\pi^+ \Lambda}$ for the unitary action.
  The notation is the same as that of Fig.~\ref{fig:5}.
}
\label{fig:7a}
\end{figure}

\begin{figure}[htpb]
\begin{center}
	{\normalsize
          \includegraphics[width=0.8\textwidth]{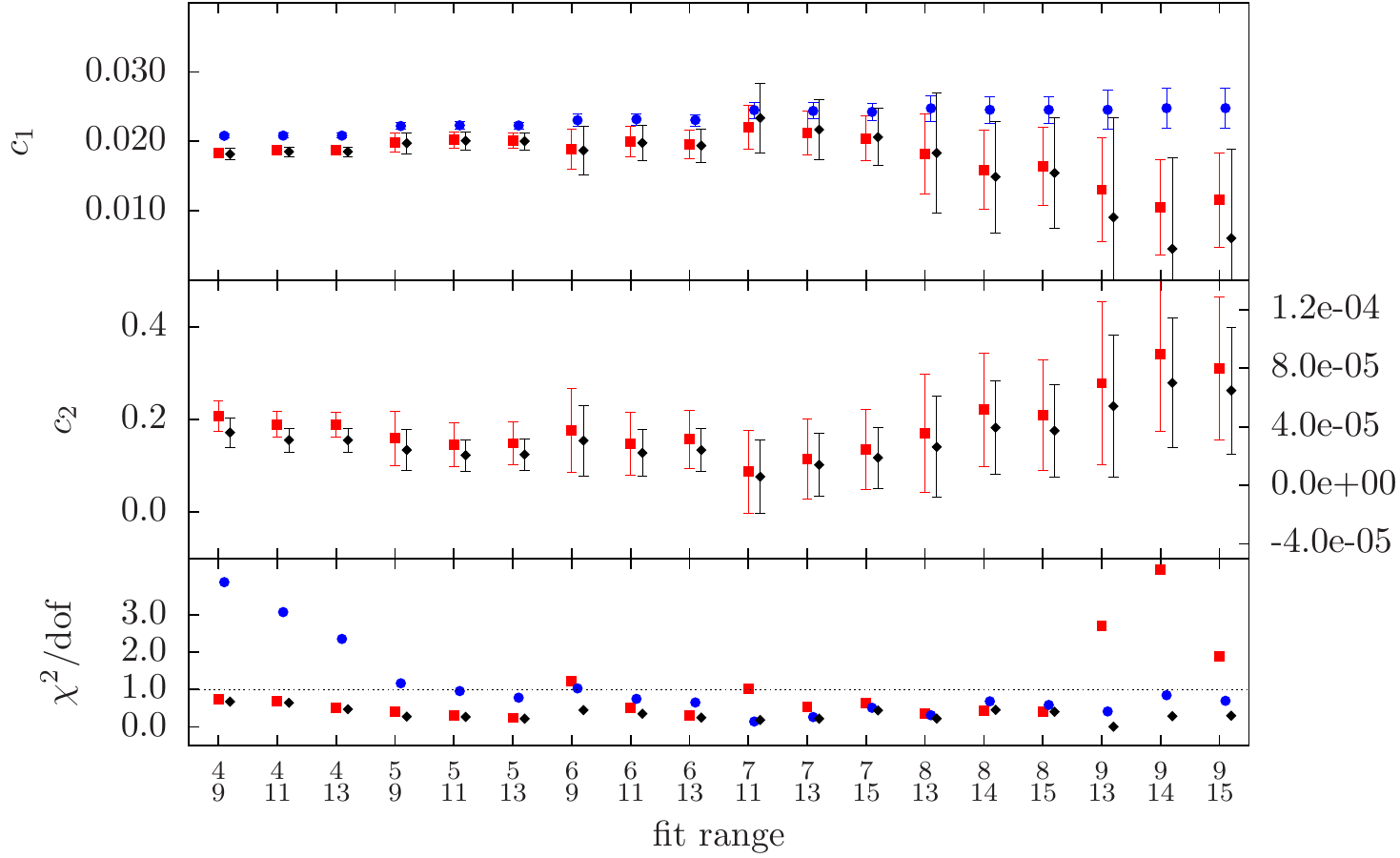}
}
\end{center}
\caption{ Same as Fig.~\reffig{fig:7a}, but for the hybrid action. }
\label{fig:7b}
\end{figure}

For the decay of the $\Sigma^*$, we follow the same approach as in the case of the $\Delta$ decay.
In Fig.~\reffig{fig:6} we show the ratio $R^{\Sigma^{*+}}_{\pi \Lambda}$ for both the unitary and hybrid cases
and Figs.~\reffig{fig:7a} and \reffig{fig:7b}
display the behavior of the three different fits when varying the fit ranges.
For the unitary action, the energy levels for $\Sigma^*$ and $\pi\,\Lambda$ are still close
and we find acceptable linear fits already starting at $t_\mathrm{min}/a = 2$.
For the hybrid case we find, that starting with $t_\mathrm{min}/a = 7$ we find a time independent value for the slope even for the
linear fit. The estimates for the slope from the cubic and hyperbolic sine fit are mutually consistent even before that.
However, the $\chisqrPerDof$ for all fit versions is acceptable starting as early as $t_\mathrm{min}/a = 5$.
A qualitative difference between the unitary and the hybrid ensembles becomes apparent when examining the ratio  $R^{\Sigma^{*+}}_{\pi \Lambda}$. When attempting a linear fit
to extract the slope, with the hybrid action, the central value for $c_1$ rises systematically, when the lower end of the fit window is moved towards larger
time-slices. This would be expected for a significant energy gap between the state excited by the $\Sigma^*$ and the $\pi\,\Sigma$ interpolating fields.
%The upward curvature then showing, that $E_{\pi\,\Lambda} > E_{\Sigma^*}$. %%%%JWN
The upward curvature then shows that $E_{\pi\,\Lambda} > E_{\Sigma^*}$.

%%% Using the same weighted analysis performed for the $\Delta$, we extract, for the unitary action,  the values
%%% \begin{align}
%%%   g^{\Sigma^*}_{\pi \Lambda}\left(\mathrm{unitary} \right) &= 18.5 \,(0.3 )\,(0.6 )
%%% \label{eq:g_sigmaps2piplambda0_unitary}\\
%%% a\Gamma^{\Sigma^*}_{\pi \Lambda}\left(\mathrm{unitary} \right) &=  0.0396 2,(15)\,(16)\,,
%%% \label{eq:width_sigmaps2piplambda0_unitary}
%%% \end{align}
%%% while for the hybrid action we find
%%% \begin{align}
%%% g^{\Sigma^*}_{\pi \Lambda}\left(\mathrm{hybrid} \right) &= 23.2 \,(0.6 )\,(2.4)
%%% \label{eq:g_sigmaps2piplambda0_hybrid}\\
%%% a\Gamma^{\Sigma^*}_{\pi \Lambda}\left(\mathrm{hybrid} \right) &= 0.0904 \,(46)\,(104)\,.
%%% \label{eq:width_sigmaps2piplambda0_hybrid}
%%% \end{align}

%%%%%%%%%%%%%%%%%%%%%%%%%%%%%%%%%%%%%%%%%%%%%%%%%%%%%%%%%%%%%%%%%%%%%%%%%%%%%%%%%%%
%   Sigma^* to pi Sigma
%%%%%%%%%%%%%%%%%%%%%%%%%%%%%%%%%%%%%%%%%%%%%%%%%%%%%%%%%%%%%%%%%%%%%%%%%%%%%%%%%%%
\subsection{$\Sigma^* \to \pi\,\Sigma$}
\begin{figure}[htpb]
\begin{center}
\includegraphics[width=0.7\textwidth]{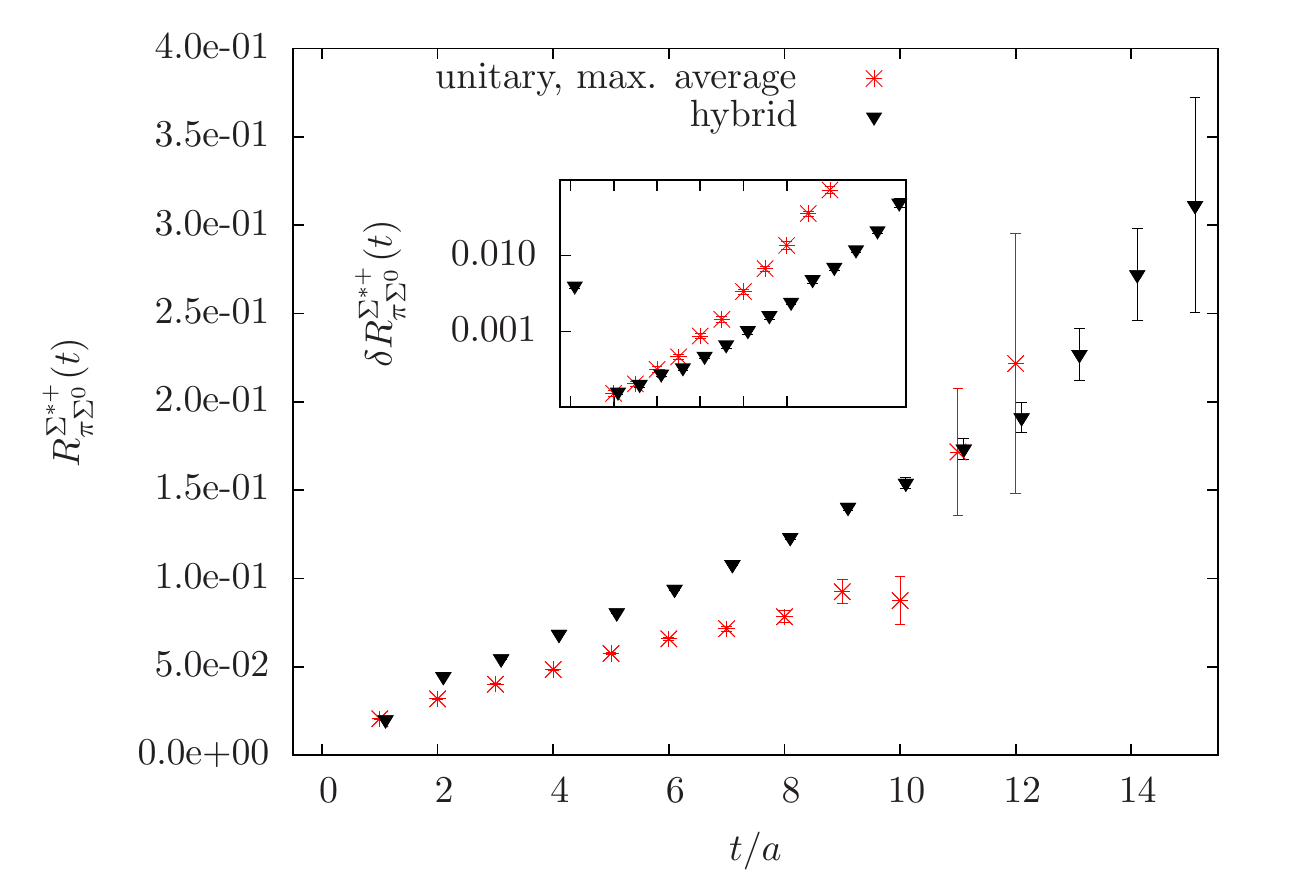}
\end{center}
\caption{
  Ratio $R^{\Sigma^{*+}}_{\pi^+ \Sigma^0}$ for the unitary and hybrid calculations.%%% s added
  The notation is the same at that in Fig. \ref{fig:4}.
}
\label{fig:8}
\end{figure}

\begin{figure}[htpb]
\begin{center}
	{\normalsize
          \includegraphics[width=0.8\textwidth]{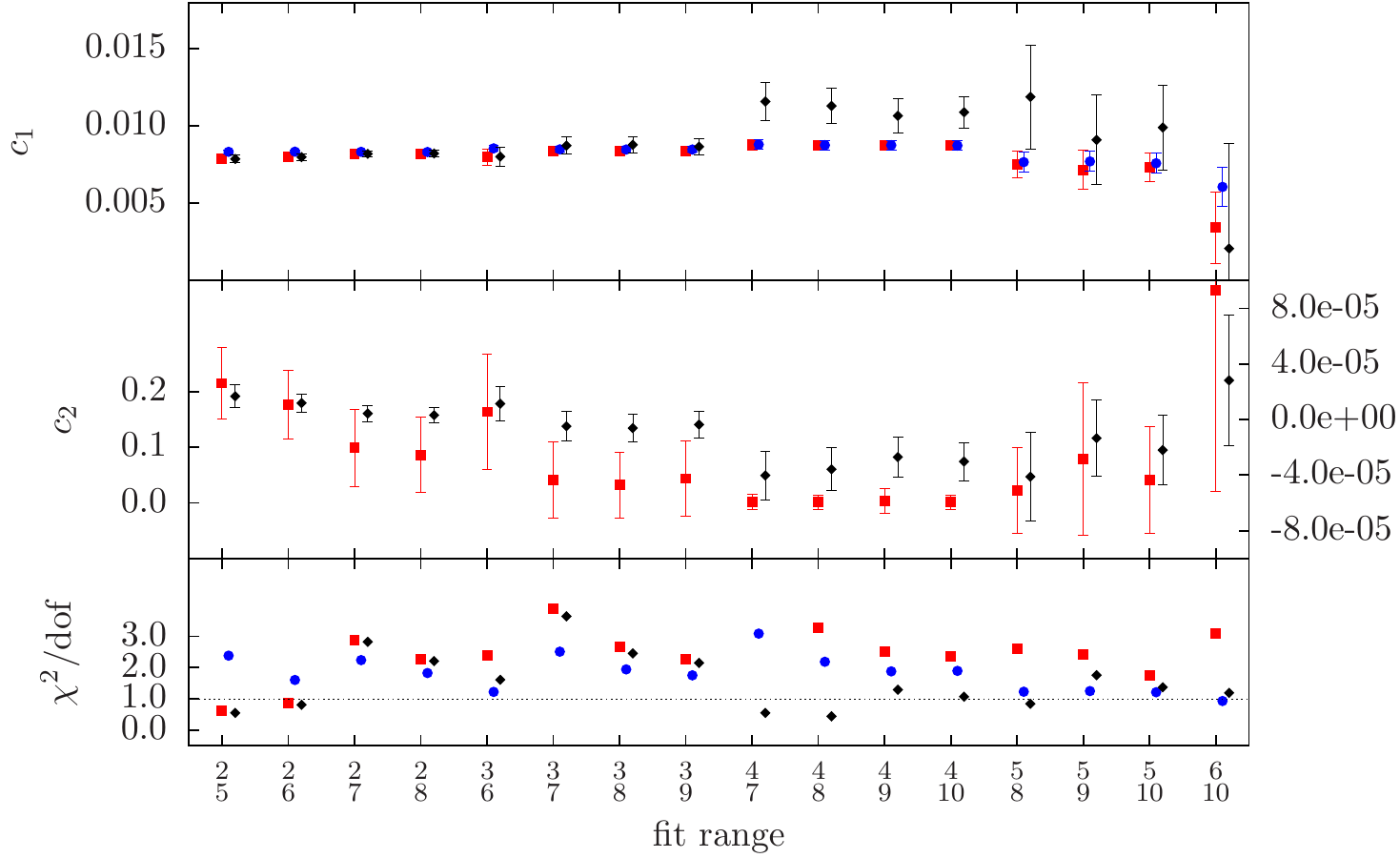}
}
\end{center}
\caption{
  Fit range dependence of $c_1,\,c_2$ for three different fits of ratio $R^{\Sigma^{*+}}_{\pi^+ \Sigma^0}$ for the unitary action.
  The notation is the same as that in Fig.~\ref{fig:5}.
}
\label{fig:9a}
\end{figure}

\begin{figure}[htpb]
\begin{center}
	{\normalsize
          \includegraphics[width=0.8\textwidth]{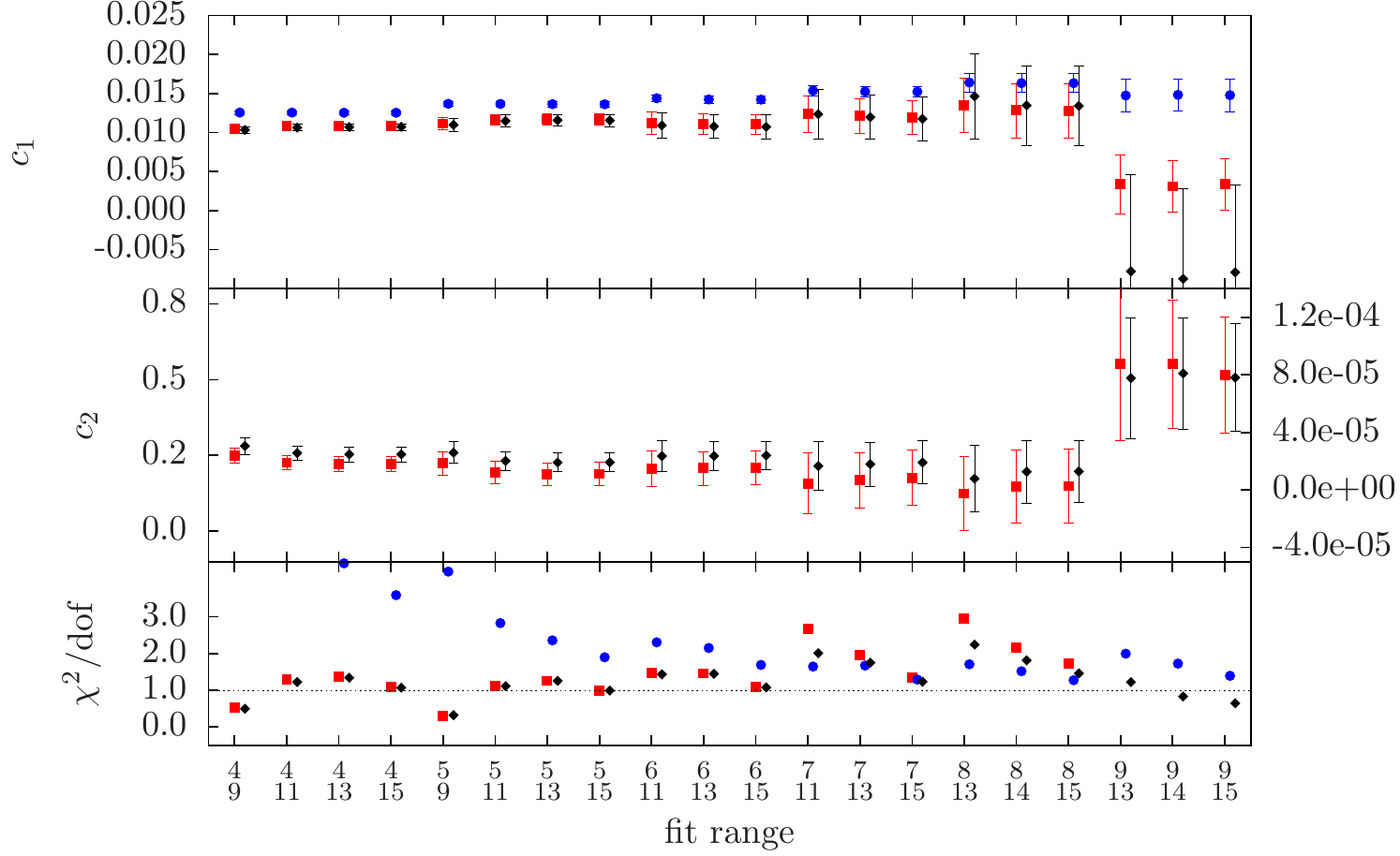}
}
\end{center}
\caption{Same as Fig.~\reffig{fig:9a}, but for the hybrid action.}
\label{fig:9b}
\end{figure}

For the transition $\Sigma^* \to \pi\,\Sigma$ we  show the results for the ratio $R^{\Sigma^*}_{\pi\,\Sigma}$ in Fig.~\reffig{fig:8}
and the results for the parameters $c_1$ and $c_2$ from a variety of our fits in Figs.~\reffig{fig:9a} for the unitary and
\reffig{fig:9b} for the hybrid calculation.

%%% The following values are extracted in the case of the unitary action:
%%% \begin{align}
%%% 	g^{\Sigma^*}_{\pi \Sigma}\left(\mathrm{unitary} \right) &= 16.1\,(0.3)\,(2.4)
%%% \label{eq:g_sigmaps2pipsigma0_unitary}\\
%%% a\Gamma^{\Sigma^*}_{\pi \Sigma}\left(\mathrm{unitary} \right) &= 0.0128 \,(06)\,(32)
%%% \label{eq:width_sigmaps2pipsigma0_unitary}
%%% \end{align}
%%% and for the hybrid action we find
%%% \begin{align}
%%% 	g^{\Sigma^*}_{\pi \Sigma}\left(\mathrm{hybrid} \right) &= 19.0 \,(0.7)\,(3.3)
%%% \label{eq:g_sigmaps2pipsigma0_hybrid}\\
%%% a\Gamma^{\Sigma^*}_{\pi \Sigma}\left(\mathrm{hybrid} \right) &= 0.0366\,(22)\,(72)
%%% \label{eq:width_sigmaps2pipsigma0_hybrid}
%%% \end{align}

%%%%%%%%%%%%%%%%%%%%%%%%%%%%%%%%%%%%%%%%%%%%%%%%%%%%%%%%%%%%%%%%%%%%%%%%%%%%%%%%%%%
%  Xi^* to Xi pi
%%%%%%%%%%%%%%%%%%%%%%%%%%%%%%%%%%%%%%%%%%%%%%%%%%%%%%%%%%%%%%%%%%%%%%%%%%%%%%%%%%%
\subsection{$\Xi^* \to \pi \Xi$}
\begin{figure}[htpb]
\begin{center}
\includegraphics[width=0.7\textwidth]{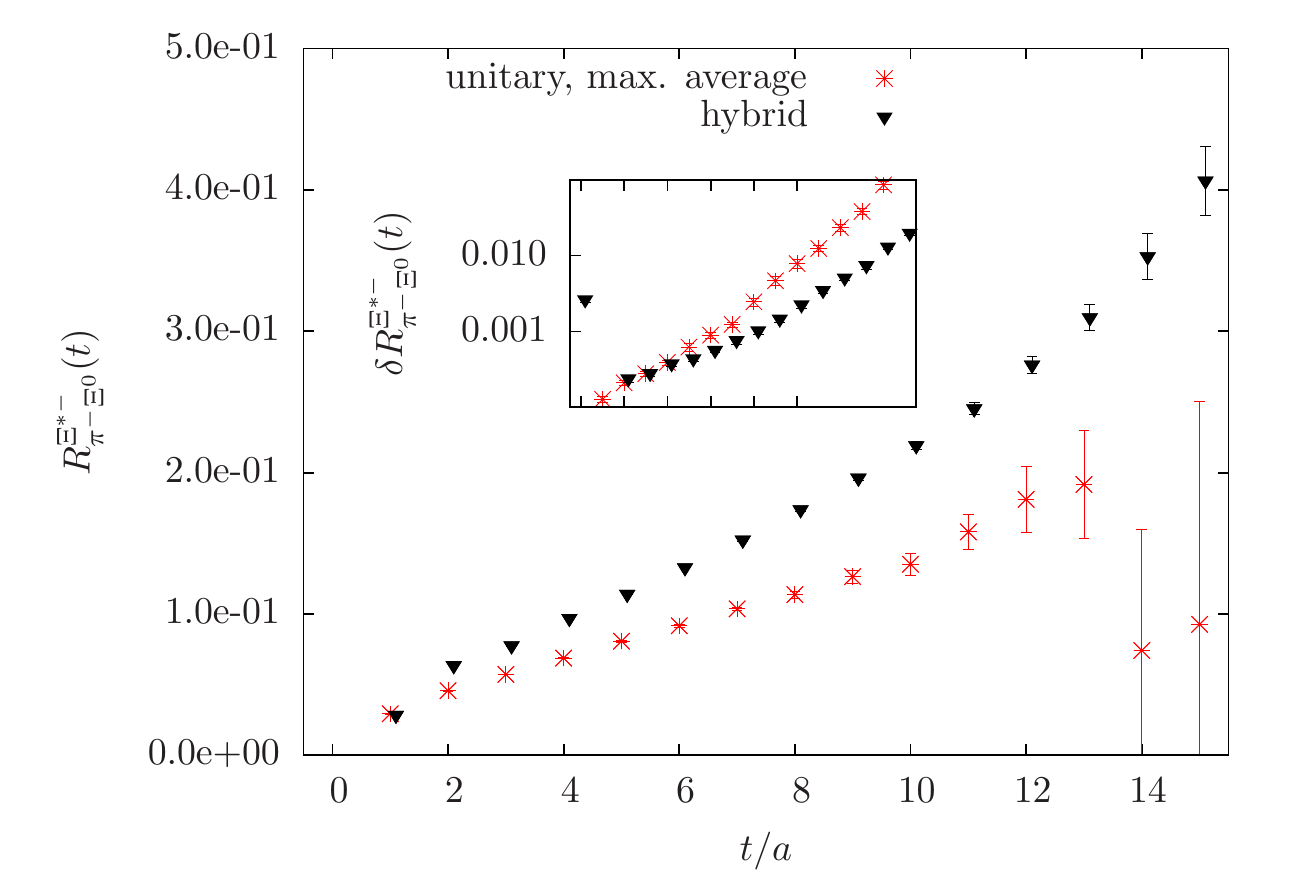}
\end{center}
\caption{
  Ratio $R^{\Xi^{*-}}_{\pi^- \Xi^0}$ for the unitary and hybrid calculations.%%% s added
  The notation is the same as that  in the Fig.~\ref{fig:4}.
}
\label{fig:10}
\end{figure}

\begin{figure}[htpb]
\begin{center}
\includegraphics[width=0.8\textwidth]{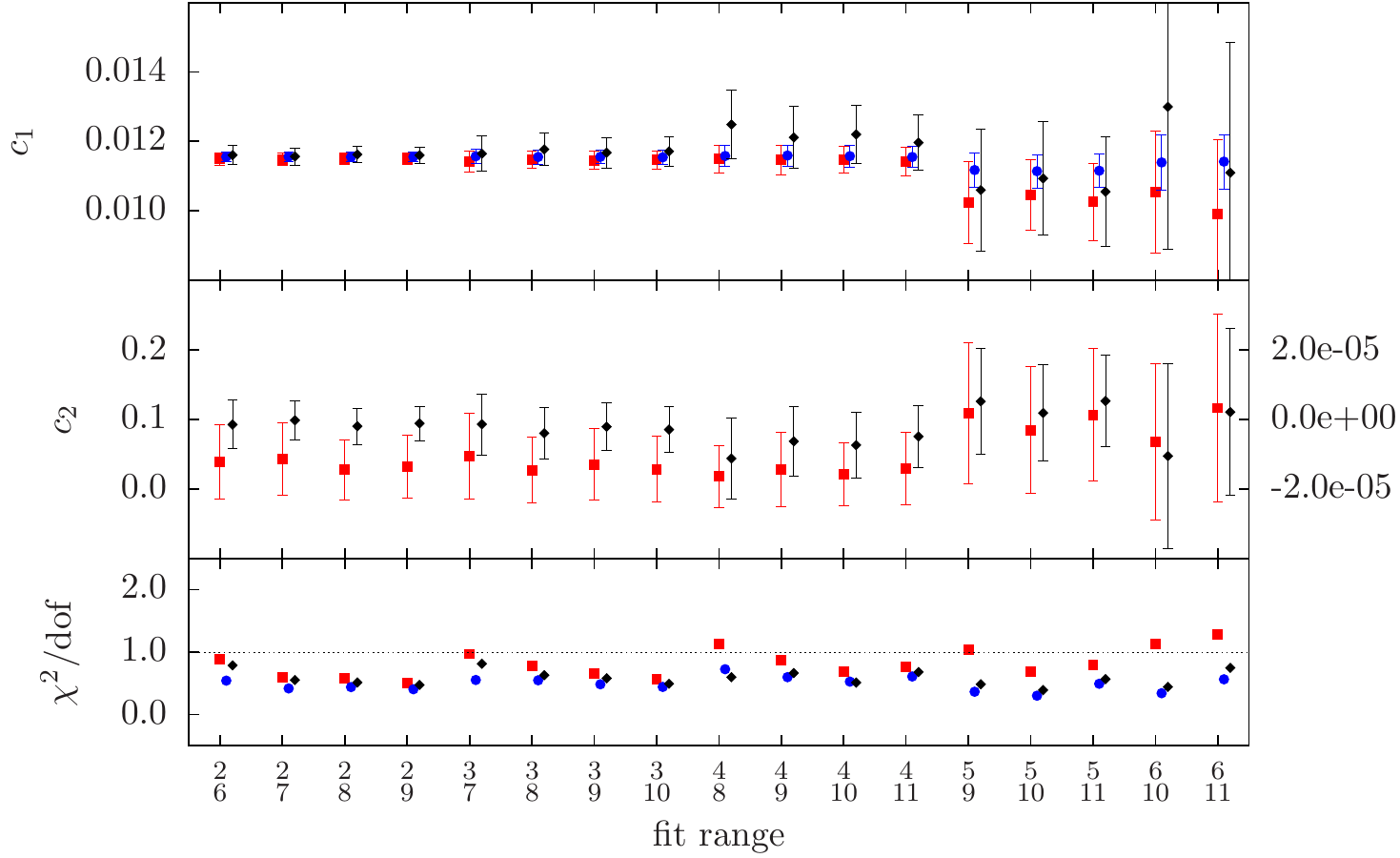}
\end{center}
\caption{
  Fit range dependence of $c_1$ and $c_2$ from the fits of ratio $R^{\Xi^{*-}}_{\pi^- \Xi^0}$ for the unitary calculation.
  The notation is the same as that  in the Fig.~\ref{fig:5}.
}
\label{fig:11a}
\end{figure}

\begin{figure}[htpb]
\begin{center}
\includegraphics[width=0.8\textwidth]{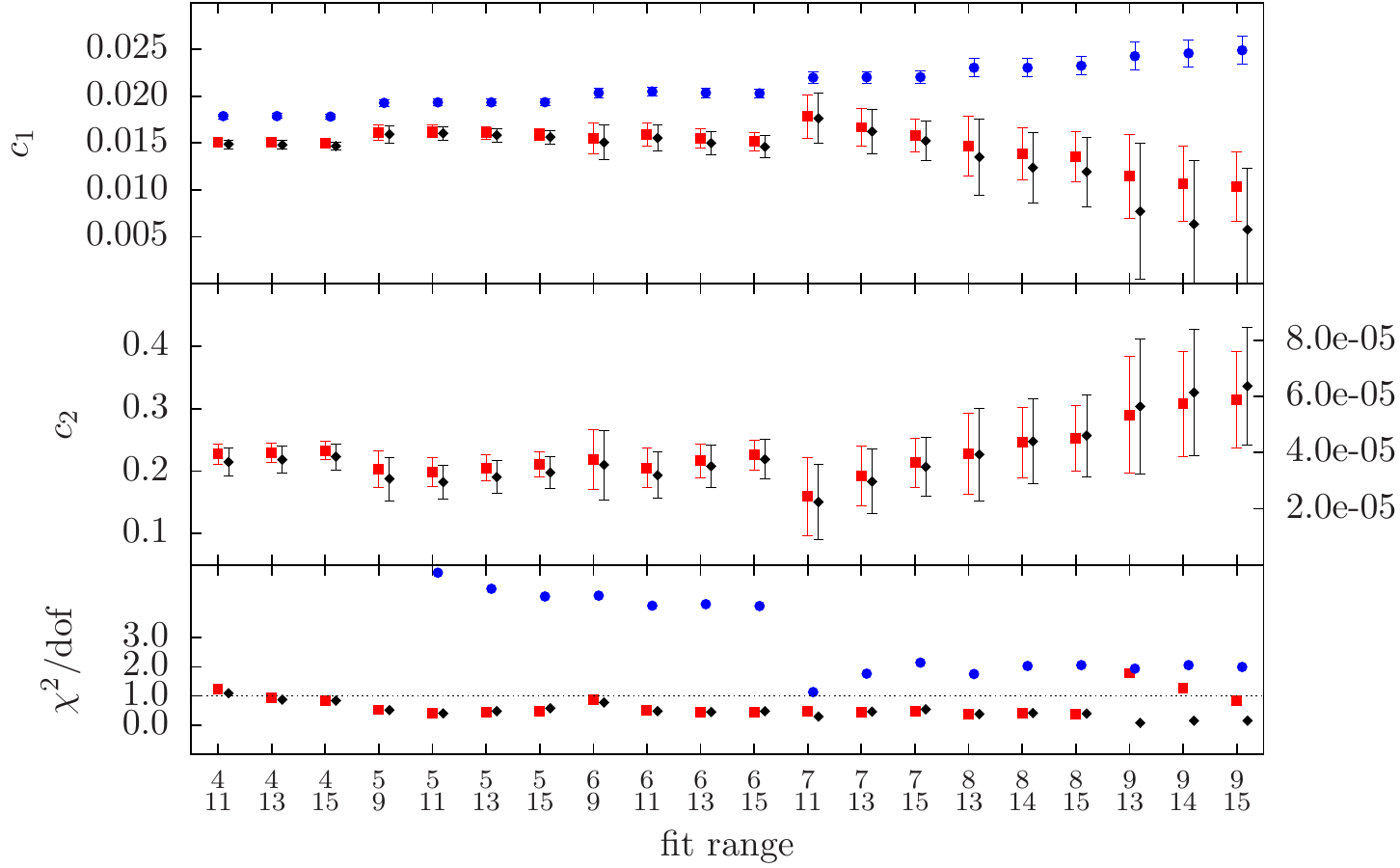}
\end{center}
\caption{Same as Fig.~\reffig{fig:11a}, but for the hybrid calculation. }
\label{fig:11b}
\end{figure}

Finally we present the results for the transition $\Xi^* \to \pi\,\Xi$ in an analogous manner
in Figs. \reffig{fig:10}, \reffig{fig:11a} and \reffig{fig:11b}.

%%% For the coupling and the width we extract the following values for the unitary action:
%%% \begin{align}
%%% 	g^{\Xi^*}_{\pi \Xi}\left(\mathrm{unitary} \right) &=  21.0 \,(0.3 )\,( 0.5)
%%% \label{eq:g_xism2pimxi0_unitary}\\
%%% a\Gamma^{\Xi^*}_{\pi \Xi}\left(\mathrm{unitary} \right) &= 0.0255 2\,(08)\,(08)
%%% \label{eq:width_xism2pimxi0_unitary}
%%% \end{align}
%%% and for the hybrid action
%%% \begin{align}
%%% 	g^{\Xi^*}_{\pi \Xi}\left(\mathrm{hybrid} \right) &= 25.6 \,(0.6)\,(6.0)
%%% \label{eq:g_xism2pimxi0_hybrid}\\
%%% a\Gamma^{\Xi^*}_{\pi \Xi}\left(\mathrm{hybrid} \right) &= 0.0792 \,(35)\,(154)
%%% \label{eq:width_xism2pimxi0_hybrid}
%%% \end{align}

%%% In Table~\reftab{tab:2} we gather our results to allow an easy comparison and convert to physical units using the values for
%%% the lattice spacing given in Table~\reftab{tab:1}.
%%% \begin{table}
%%% \begin{center}
%%% \input{results_physical_units_table.tex}
%%% \end{center}
%%% \caption{
%%%   For each decay process given in the first column,
%%%   we give the $MB\>B^*$ coupling constant $g_{MB}^{B^*}$ (second column)
%%%   for the DWF and hybrid actions, and corresponding width $\Gamma_{MB}^{B*}$ in $\mev$.
%%%   The uncertainties in brackets are statistical and systematic as given in
%%%   subsection \ref{subsec:delta_to_pi_n}.
%%% }
%%% \label{tab:2}
%%% \end{table}
We gather our results for the coupling and widths in Tables~\reftab{tab:results_coupling} and \reftab{tab:results_width} below.
To allow for an easy comparison we convert the decay widths to physical units using the values for
the lattice spacing given in Table~\reftab{tab:1}.

The results for the process $\Delta \leftrightarrow \pi\,N$ with the hybrid action 
differ slightly from our previous investigation~\cite{Alexandrou:2013ata,Alexandrou:2014qka},
since we updated them using the weighted average for the distribution of fits.

Utilizing the expressions of Eqs.~\refeq{eq:coupling_formula} and \refeq{eq:matelem_formula}
we estimate the coupling, which is independent of the isospin combination
of in and out state, while the width is for specific combinations of in and out states.
For this reason, in the table we distinguish explicitly the isospin
dependence of the width by giving
the electromagnetic charges of the interpolating fields as superscripts.

%%%%%%%%%%%%%%%%%%%%%%%%%%%%%%%%%%%%%%%%%%%%%%%%%%%%%%%%%%%%%%%%%%%%
% results for study of AMA; comparison with previous result
%%%%%%%%%%%%%%%%%%%%%%%%%%%%%%%%%%%%%%%%%%%%%%%%%%%%%%%%%%%%%%%%%%%%
\subsection{Improved precision with AMA for $R^\Delta_{\pi\,N}$}
In order to assess the potential of using all-mode-averaging to improve the accuracy of our computations,
we apply all-mode-averaging~\cite{Blum:2012uh} on a subset of 89 (out of 254) configurations and specifically look
at the case of $\Delta \to \pi\,N$. In addition to the correlation functions, which had been obtained at high solver precision during
the production of quark propagators, a corresponding data set at low solver precision was produced with 16 random shifts of the original
spatial source position two-point correlation functions $C_{\Delta - \Delta}$, $C_{\pi - \pi}$ and $C_{N - N}$
for each of the four preset source time-slices independently.
The measurements for $C_{\Delta - \pi N}$ are  done coherently with a single inversion after inserting sequential sources at the four time-slices.

\begin{figure}[htpb]
\begin{center}
\includegraphics[width=0.7\textwidth]{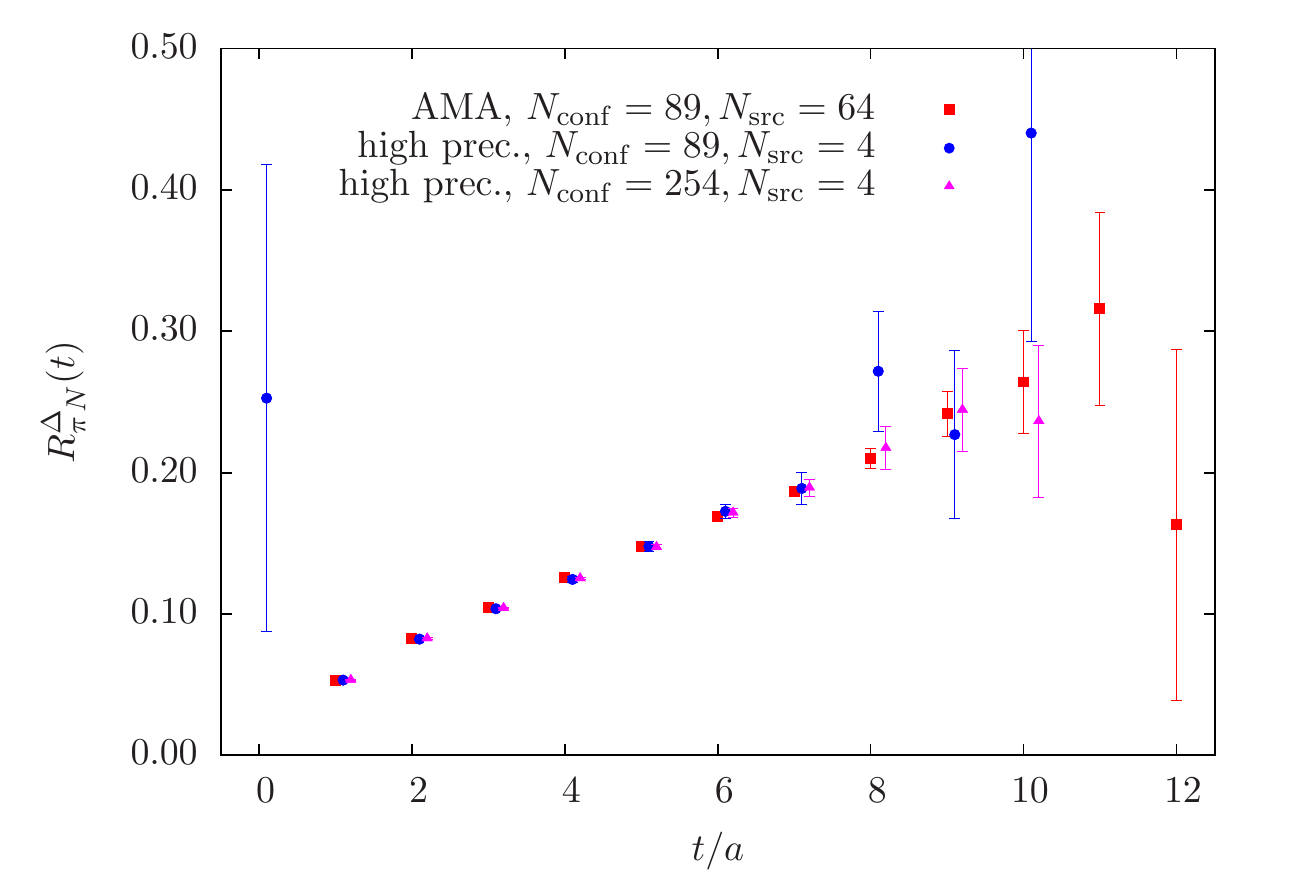}
\end{center}
\caption{
  Comparison of results on the ratio $R^\Delta_{\pi\,N}$ with (red squares) and without using all-mode-averaging (blue circles and magenta triangles).
}
\label{fig:12a}
\begin{center}
\includegraphics[width=0.7\textwidth]{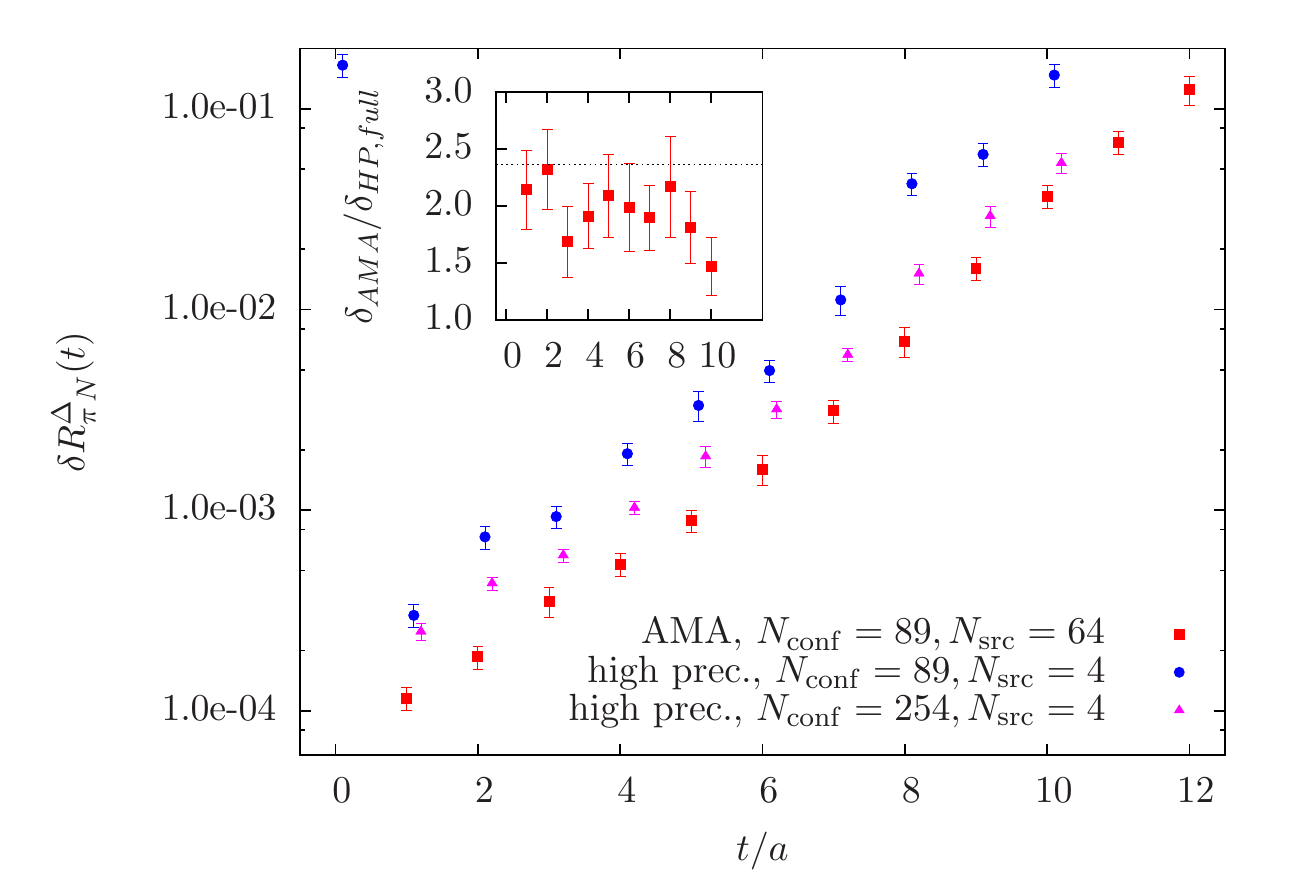}
\end{center}
\caption{
  Comparison of the statistical uncertainties of $R^\Delta_{\pi\,N}$ with and without using all-mode-averaging; the detail plot
  shows the ratio of the uncertainty obtained by using AMA over the that from the original run with 254 configurations and 4 source time-slices
  per configuration; the horizontal line marks the ratio of uncertainties expected for ideal error scaling.
}
\label{fig:12b}
\end{figure}

We show a comparison of the estimates for the ratio $R^\Delta_{\pi\,N}$ and its statistical uncertainty in Figs.
\reffig{fig:12a} and \reffig{fig:12b}. We find full consistency of the data for the ratio from both the AMA simulation and the
original production run. Moreover, the uncertainty is reduced by a factor around two across the relevant time-slices $1\le t/a \le 10$.
From an ideal scaling of the error we expect reduction of the statistical uncertainty by a factor of
$\sqrt{ \left( 89\times 64 \right) / \left( 254 \times 4 \right) } \approx 2.4$ and % this 
the observed behavior is consistent with % the 
this expectation.

We determine the coupling and width based on the AMA data set in the way previously outlined
and give the results for comparison in Table~\ref{tab:AMA}.
\begin{table}
\begin{center}
\begin{tabular}{c|cc}
  & $g^{\Delta}_{\pi\,N}$ & $a\Gamma^\Delta_{\pi\,N}$ \\
\hline
 AMA      & $24.2   \,(0.3) \,(1.0)$  & $0.0846 \,(23)  \,(22)$  \\
 HP, full & $23.7   \,(0.7) \,(1.1)$  & $0.0868 \,(57)  \,(45)$ \\
\hline
\hline
\end{tabular}
\caption{
  The results on the $\Delta-\pi N$ coupling constant  and width using AMA are shown in the first row,
  while results extracted without AMA are included in the second row.
}
\label{tab:AMA}
\end{center}
\end{table}
%%%%%%%%%%%%%%%JWN
%The comparison of the uncertainties in the previous table shows, that a in the final results an improvement of both the statistical and systematic
%uncertainty by a factor of 2 is maintained.
The statistical errors show the expected improvement by a factor of approximately 2.4.

\section{Discussion}
\label{sec:Discussion}
In order to make a direct comparison with the experimental values, we provide 
in Table~\ref{tab:experiment} the data taken from the Particle Data Group~\cite{Beringer:1900zz} for the 
relevant baryon and meson masses, the full widths, branching ratios and relative momentum $k$ of the asymptotic final 
meson and baryon state from the decay in the center-of-mass frame. The coupling constant
is then derived according to the tree-level decay process using the expression in Eq.~(\ref{eq:width_lo_eft}) and the experimental value of the width
as an input.

We compare these values for the coupling constants to the results of our calculation in Table~\reftab{tab:results_coupling}.
The analogous comparison for the decay widths in physical units is shown in Table~\reftab{tab:results_width}.
%%% {\bf Dina:}
We would like to stress that although we show the results for the hybrid and unitary calculation side-by-side in the tables, one should be careful
in drawing strong conlcusions since
the conditions for the applicability of the transfer
matrix method are fulfilled to different degree in the two cases.
In particular, the energy matching is very different in the two cases and for a direct comparison one would need to have kinematics where the energy gap is similar.

\begin{table}
\begin{center}
\begin{tabular}{l|l|ccc|cc|c}
$B^*$ & $M B$ & $m_{B^*} / \mev$ & $m_M / \mev$ & $m_B / \mev$ & $\Gamma_\mathrm{full} / \mev$ & $\Gamma^{B^*}_{MB} / \Gamma_\mathrm{full} $ & $p / \mev$ \\
\hline
$\Delta$   & $\pi\,N$       & $1232\,(1)$     & $139.57018\,(35)$ & $938.272013\,(23)$ & $118\,(2)$   & $1.$          & $227$ \\
$\Sigma^*$ & $\pi\,\Lambda$ & $1382.80\,(35)$ &                   & $1115.683\,(6)$    & $36.0\,(7)$  & $0.870\,(15)$ & $205$ \\
$\Sigma^*$ & $\pi\,\Sigma$  &                 &                   & $1192.642\,(24)$   &              & $0.117\,(15)$ & $120$ \\
$\Xi^*$    & $\pi\,\Xi$     & $1535.0\,(6)$   &                   & $1314.86\,(20)$    & $9.9\,(1.9)$ & $1.$          & $158$ \\
\hline
\hline
\end{tabular}
\end{center}
\caption{The physical values of masses $m_{B^*}$, $m_B$ and $m_M$, full widths $\Gamma_\mathrm{full}$, branching ratios $\Gamma^{B^*}_{M B}$
  and relative momentum $p$ for the 2-hadron state
  for the resonances studied in this work as given by the Particle Data Group~\cite{Beringer:1900zz}.
%%% The coupling constant $g_{MB}^{B^*}$ is derived
%%% using the experimental value of the width and the effective field theory result to leading  order.
%%% The errors on the values quoted  in this table for the coupling are calculated by naive error propagation without taking into account correlations among
%%% the input quantities.
}
\label{tab:experiment}
\end{table}

\begin{table}
  \centering
\begin{tabular}{l|cc|c}
process & unitary & hybrid & PDG \\
	\hline
%$\Delta^{++} \leftrightarrow \pi^+\,N^+$      & $ 23.7\,( 0.7)\,( 2.0)$ & $ 26.7\,( 0.6)\,( 2.3)$ & $29.4\,(0.3)$ \\
%$\Sigma^{*+} \leftrightarrow \pi^+\,\Lambda$  & $ 18.5\,( 0.3)\,( 0.6)$ & $ 23.2\,( 0.6)\,( 2.4)$ & $20.4\,(0.3)$ \\
%$\Sigma^{*+} \leftrightarrow \pi^+\,\Sigma^0$ & $ 16.1\,( 0.3)\,( 2.4)$ & $ 19.0\,( 0.7)\,( 3.3)$ & $17.3\,(1.1)$ \\
%$\Xi^{*-} \leftrightarrow \pi^-\,\Xi^0$       & $ 21.0\,( 0.3)\,( 0.5)$ & $ 25.6\,( 0.6)\,( 6.0)$ & $19.4\,(1.9)$ \\
$\Delta^{++} \leftrightarrow \pi^+\,N^+$      & $ 23.7\,( 0.7)\,( 1.1)$ & $ 26.7\,( 0.6)\,( 1.4)$ & $29.4\,(0.3)$ \\
$\Sigma^{*+} \leftrightarrow \pi^+\,\Lambda$  & $ 18.5\,( 0.3)\,( 0.5)$ & $ 23.2\,( 0.6)\,( 0.8)$ & $20.4\,(0.3)$ \\
$\Sigma^{*+} \leftrightarrow \pi^+\,\Sigma^0$ & $ 16.1\,( 0.3)\,( 1.9)$ & $ 19.0\,( 0.7)\,( 2.9)$ & $17.3\,(1.1)$ \\
$\Xi^{*-} \leftrightarrow \pi^-\,\Xi^0$       & $ 21.0\,( 0.3)\,( 0.3)$ & $ 25.6\,( 0.6)\,( 4.3)$ & $19.4\,(1.9)$ \\
\hline
\hline
\end{tabular}
  \caption{
    Results for the couplings $g^{B^*}_{MB}$.
    For each decay process given in the first column,
    we give the coupling constant $g_{MB}^{B^*}$
    %%%%%%%%%%%%%%%JWN
%    for the DWF (second column) and hybrid actions (third column). 
 for the  unitary DWF ensemble with $m_\pi$ = 180 MeV (second column) and hybrid ensemble with $m_\pi$ = 350 MeV (third column). 
    The fourth column shows
    the value of the coupling at leading order effective field theory using input from the PDG.
    The uncertainties in brackets are statistical and systematic as given in
    subsection \ref{subsec:delta_to_pi_n}.
}
  \label{tab:results_coupling}
\end{table}
 
\begin{table}
  \centering
\begin{tabular}{l|cc|c}
process & unitary & hybrid & PDG \\
\hline
%$\Delta^{++} \leftrightarrow \pi^+\,N^+$      & $ 119.4\,( 7.9)\,( 8.1)$ & $238.5\,( 12.2)\,( 19.8)$ & $118\,(2)$ \\
%$\Sigma^{*+} \leftrightarrow \pi^+\,\Lambda$  & $  54.5\,( 2.1)\,( 2.2)$ & $ 143.9\,( 7.4)\,( 16.5)$ & $31.3\,(8)$ \\
%$\Sigma^{*+} \leftrightarrow \pi^+\,\Sigma^0$ & $  17.6\,( 0.8)\,( 4.5)$ & $  58.3\,( 3.4)\,( 11.5)$ & $2.1\,(3)$ \\
%$\Xi^{*-} \leftrightarrow \pi^-\,\Xi^0$       & $  35.1\,( 1.1)\,( 1.1)$ & $ 126.0\,( 5.6)\,( 24.6)$ & $6.6\,(1.2)$ \\
$\Delta^{++} \leftrightarrow \pi^+\,N^+$      & $ 119.4\,( 7.9)\,( 4.5)$ & $238.5\,( 12.2)\,( 16.2)$ & $118\,(2)$ \\
$\Sigma^{*+} \leftrightarrow \pi^+\,\Lambda$  & $  54.5\,( 2.1)\,( 1.3)$ & $ 143.9\,( 7.4)\,(  6.1)$ & $31.3\,(8)$ \\
$\Sigma^{*+} \leftrightarrow \pi^+\,\Sigma^0$ & $  17.6\,( 0.8)\,( 2.1)$ & $  58.3\,( 3.4)\,(  6.8)$ & $4.2\,(5)$ \\
$\Xi^{*-} \leftrightarrow \pi^-\,\Xi^0$       & $  35.1\,( 1.1)\,( 0.4)$ & $ 126.0\,( 5.6)\,( 18.5)$ & $9.9\,(1.9)$ \\
\hline
\hline
\end{tabular}
  \caption{Results for the decay
  %%%%%%%%%%%%%%%%%%%%%%%%%%%%JWN 
%   widths. The meaning of the columns is analogous to Table~\reftab{tab:results_coupling} with decay process (first column), results from
%    unitary DWF (second column) and hybrid actions (third column) and the PDG value (fourth column).}
   widths in MeV. The meaning of the columns is analogous to Table~\reftab{tab:results_coupling} with the decay process (first column), results from the
    unitary DWF  ensemble with $m_\pi$ = 180 MeV (second column),   results from the
    hybrid  ensemble with $m_\pi$ = 350 MeV (third column) and the PDG value (fourth column).}
  \label{tab:results_width}
\end{table}

We find that our lattice QCD values for the couplings are in good agreement with the PDG-derived values for all decays
for both the unitary and the hybrid action with a tendency of higher values
for the latter case. This observed level of agreement is remarkable,
given that with the unitary and hybrid action we simulate at pion mass $180\mev$ and $350\mev$, respectively,
and on coarse lattices.

For the width itself, on the other hand, we only find agreement for $\Gamma^\Delta_{\pi\,N}$ with the unitary action.
This may be expected since it is only for this case
 that the energies 
of the states $B^*$ and $M\,B$ are degenerate and therefore this case is the closest to the threshold situation where the conditions of our approach are best fulfilled.
Table~\ref{tab:experiment} shows in the right-most column the momentum in the center-of-mass frame for the fields $M$ and $B$ for the individual decays.
On the lattice, this momentum is of course fixed to $k = 2\pi/L$ in lattice units or $k(\mathrm{unitary}) \approx 270\mev$ for the unitary
action and $k(\mathrm{hybrid}) \approx 357 \mev$ for the hybrid one. Thus, in addition to matching the energies of the resonance and the decay channel,
one has another constraint, namely,  a fixed center-of-mass momentum in these transition processes, which deviates
from the physical situation by a process-dependent amount. In general, we have that with the hybrid action, the lattice momentum is 1.5 to 3 times larger
than its value in the continuum infinite volume limit. With the unitary action the violation is less severe, and the closest to the physical situation  is the one corresponding to decay of the $\Delta$.

%%% The coupling constant, being a dimensionless quantity, is thus likely to have a weaker dependence on these violations stemming from dimensionful
%%% parameters like particle masses and energies, and momenta.

Assuming a finite volume we can check that the density of states derived from the
lattice values of the masses and the momentum approaches closer to the value of the density of states derived with their continuum counterparts when going from
$\pi\,N$ to $\pi\,\Lambda / \pi\,\Sigma$ to $\pi\,\Xi$. This is to be expected, since the strange quark mass is tuned closer to its physical value
than the light quark mass and $\Lambda/\Sigma$ and $\Xi$ are have strangeness $-1$ and $-2$, respectively.

%%% {\bf Dina: I did not fully understand the point here so I modified a bit. Please check}
The dependence of the coupling and decay width on the meson and baryon masses, momentum and the parameters of the lattice simulation show a large 
disparity 
 reflected in the different levels of agreement in Tables~\reftab{tab:results_coupling} and \reftab{tab:results_width}.
Partly this is explained by the additional condition of having to match the
center of mass momentum for extracting the width in the decay process.
One would need to study the dependence of the momentum further in order to understand
the different level of agreement 
between the case of the coupling and that of the width.

\section{Conclusions and Outlook}
\label{sec:Conclusions}
The coupling constants $g_{\pi N}^\Delta$, $g_{\pi \Lambda}^{\Sigma^*}$, $g_{\pi \Sigma}^{\Sigma^*}$ and $g_{\pi \Xi}^{\Xi^*}$ are evaluated using two ensembles of
dynamical fermion gauge configurations with pion mass 350~MeV  and 180~MeV.
In both cases, domain wall valence quarks are used. The gauge configurations for the ensemble with the heavier mass were produced using $N_f=2+1$ staggered sea quarks and thus our analysis is done with a hybrid action, while those with the lighter pion mass were produced using $N_f=2+1$ domain wall sea quarks so the action is unitary.
The kinematical conditions are best satisfied for the unitary action for all four decays,
with the $\Delta$-decay being closest to the physical situation.
Comparing the values of the  coupling constants obtained for these two ensembles, we find that they are about 10\% smaller for the ensemble with $180\mev$ pions as compared to their values for the ensemble with 350 MeV pions.  Given that the pion mass is about half as compared to the hybrid ensemble, we conclude that the pion mass dependence is rather weak and thus the values obtained using $180\mev$ pions should be close to their values at the physical point, which is indeed what we observe.
In order to extract the width, one needs to make further assumptions, some of which are not well satisfied. For example, the
energy in the center of mass frame  on the lattice is different from the one
   in the infinite volume limit.
The case for which these energies best match is for the $\Delta$-decay where indeed we find an agreement with the experimental value.
This demonstrates that the methodology works when the physical kinematical conditions are approximately  satisfied on the lattice.

To explore the applicability of AMA
on reducing  the statistical uncertainty of the ratio and consecutively of the extracted slope, we consider the all-mode-averaging technique for
the case of the transition $\Delta \leftrightarrow \pi\,N$. Adding further correlation functions with randomly shifted source positions at
low precision for a subset of the gauge ensemble, we increase the available statistics by a factor of approximately $5.6$. The ideally expected
reduction of the statistical uncertainty is thus by a factor of $2.4$. We observe an improvement of approximately a factor of $2$ on our final, derived quantities, which is satisfactory.
 The solver precision for the low-precision inversions of the domain-wall Dirac operator is tuned to a compromise value that, on the one hand, yields
sufficiently
high statistical correlation for the two-point  functions for both high and low precision inversions  to ensure a good scaling of the statistical uncertainty  with the number of low-precision inversions, and on the other hand, to keep the ratio of cost for a low- to a high-precision propagator as small as possible, which in our case  turns out to be 1:5.

Fully exploiting the potential of further reduction of the uncertainty of the slope bears the interesting prospect of becoming sensitive to
contributions from excited states and next-to-leading order terms. This would be of particular importance in a more comprehensive, combined analysis of
several decay channels and vital for an attempt to tackle the quark-connected diagrams,
the calculation of which is beyond the scope of this work.
Notwithstanding these future prospects,  our current analysis shows, that for the time being, the major source of systematic uncertainty stems from
the lattice kinematical setup  rather than statistics.

Given the good agreement of our lattice QCD results with the experimental values
for the coupling constants and for the width when the kinematical constraints are satisfied, 
we plan to apply the method to study other baryon decays such as the decay of baryons in the negative parity channel and
decays of baryon of higher spin.

In the future we are also planning to address some of the deficiencies of the method
connected to the kinematical conditions by considering moving frames.
The decays considered here can be the test-bed for these extensions.

\section*{Acknowledgments}
%%%%%%%%%%%%%%%%%%%%JWN
%This research was in part supported by the Research Executive Agency of the
%European Union under Grant Agreement number PITN-GA-2009-238353 (ITN STRONGnet)
%and in part by the DOE Office of Nuclear Physics under grant DE-SC0011090.
This research was in part supported by the Research Executive Agency of the
European Union under Grant Agreement number PITN-GA-2009-238353 (ITN STRONGnet),
the U.S. Department of Energy Office of Nuclear Physics under grants DE-SC0011090, ER41888, and 
DE-AC02-05CH11231, and the RIKEN Foreign Postdoctoral Researcher Program.
The computing resources were provided by the National Energy Research Scientific
Computing Center supported by the Office
of Science of the DOE under Contract No. DE-AC02-05CH11231,
the J\"ulich Supercomputing Center, awarded under the PRACE EU FP7 project 2011040546
and by the Cy-Tera machine at the Cyprus Institute supported in
part by the Cyprus Research Promotion Foundation  under contract
NEA Y$\Pi$O$\Delta$OMH/$\Sigma$TPATH/0308/31.
The multi-GPU domain wall inverter code~\cite{Strelchenko:2012aa} is based on the QUDA library~\cite{Clark:2009wm,Babich:2011np}
and its development has been supported by PRACE grants RI-211528 and FP7-261557.
%The authors are grateful to Andrew Pochinsky and Sergey Syritsyn for their technical support on the QLUA interpreter language.

%%%\appendix

\bibliographystyle{h-physrev}
\bibliography{bibliography}

\end{document}